\documentclass[sigconf,nonacm,9pt]{acmart}

\usepackage[normalem]{ulem}

\usepackage[]{hyperref}

\usepackage{xspace}
\usepackage{enumitem}
\usepackage{graphicx}
\usepackage{xcolor}
\usepackage{cleveref}
\usepackage{array,tabularx}
\crefformat{section}{\S#2#1#3}
\crefformat{subsection}{\S#2#1#3}
\crefformat{subsubsection}{\S#2#1#3}
\usepackage[draft]{minted}
\usepackage{graphicx}
\usepackage{subcaption}
\newcommand{\SYSNAME}{{Hydra}\xspace}
\newcommand{\SYSNAMEPRIME}{{Hydra*}\xspace}

\newenvironment{conditions*}
  {\par\vspace{\abovedisplayskip}\noindent
   \tabularx{\columnwidth}{>{$}l<{$} @{\ : } >{\raggedright\arraybackslash}X}}
  {\endtabularx\par\vspace{\belowdisplayskip}}

\makeatletter
  \def\PYGdefault@reset{\let\PYGdefault@it=\relax \let\PYGdefault@bf=\relax%
      \let\PYGdefault@ul=\relax \let\PYGdefault@tc=\relax%
      \let\PYGdefault@bc=\relax \let\PYGdefault@ff=\relax}
  \def\PYGdefault@tok#1{\csname PYGdefault@tok@#1\endcsname}
  \def\PYGdefault@toks#1+{\ifx\relax#1\empty\else%
      \PYGdefault@tok{#1}\expandafter\PYGdefault@toks\fi}
  \def\PYGdefault@do#1{\PYGdefault@bc{\PYGdefault@tc{\PYGdefault@ul{%
      \PYGdefault@it{\PYGdefault@bf{\PYGdefault@ff{#1}}}}}}}
  \def\PYGdefault#1#2{\PYGdefault@reset\PYGdefault@toks#1+\relax+\PYGdefault@do{#2}}
  
  \@namedef{PYGdefault@tok@w}{\def\PYGdefault@tc##1{\textcolor[rgb]{0.73,0.73,0.73}{##1}}}
  \@namedef{PYGdefault@tok@c}{\let\PYGdefault@it=\textit\def\PYGdefault@tc##1{\textcolor[rgb]{0.25,0.50,0.50}{##1}}}
  \@namedef{PYGdefault@tok@cp}{\def\PYGdefault@tc##1{\textcolor[rgb]{0.74,0.48,0.00}{##1}}}
  \@namedef{PYGdefault@tok@k}{\let\PYGdefault@bf=\textbf\def\PYGdefault@tc##1{\textcolor[rgb]{0.00,0.50,0.00}{##1}}}
  \@namedef{PYGdefault@tok@kp}{\def\PYGdefault@tc##1{\textcolor[rgb]{0.00,0.50,0.00}{##1}}}
  \@namedef{PYGdefault@tok@kt}{\def\PYGdefault@tc##1{\textcolor[rgb]{0.69,0.00,0.25}{##1}}}
  \@namedef{PYGdefault@tok@o}{\def\PYGdefault@tc##1{\textcolor[rgb]{0.40,0.40,0.40}{##1}}}
  \@namedef{PYGdefault@tok@ow}{\let\PYGdefault@bf=\textbf\def\PYGdefault@tc##1{\textcolor[rgb]{0.67,0.13,1.00}{##1}}}
  \@namedef{PYGdefault@tok@nb}{\def\PYGdefault@tc##1{\textcolor[rgb]{0.00,0.50,0.00}{##1}}}
  \@namedef{PYGdefault@tok@nf}{\def\PYGdefault@tc##1{\textcolor[rgb]{0.00,0.00,1.00}{##1}}}
  \@namedef{PYGdefault@tok@nc}{\let\PYGdefault@bf=\textbf\def\PYGdefault@tc##1{\textcolor[rgb]{0.00,0.00,1.00}{##1}}}
  \@namedef{PYGdefault@tok@nn}{\let\PYGdefault@bf=\textbf\def\PYGdefault@tc##1{\textcolor[rgb]{0.00,0.00,1.00}{##1}}}
  \@namedef{PYGdefault@tok@ne}{\let\PYGdefault@bf=\textbf\def\PYGdefault@tc##1{\textcolor[rgb]{0.82,0.25,0.23}{##1}}}
  \@namedef{PYGdefault@tok@nv}{\def\PYGdefault@tc##1{\textcolor[rgb]{0.10,0.09,0.49}{##1}}}
  \@namedef{PYGdefault@tok@no}{\def\PYGdefault@tc##1{\textcolor[rgb]{0.53,0.00,0.00}{##1}}}
  \@namedef{PYGdefault@tok@nl}{\def\PYGdefault@tc##1{\textcolor[rgb]{0.63,0.63,0.00}{##1}}}
  \@namedef{PYGdefault@tok@ni}{\let\PYGdefault@bf=\textbf\def\PYGdefault@tc##1{\textcolor[rgb]{0.60,0.60,0.60}{##1}}}
  \@namedef{PYGdefault@tok@na}{\def\PYGdefault@tc##1{\textcolor[rgb]{0.49,0.56,0.16}{##1}}}
  \@namedef{PYGdefault@tok@nt}{\let\PYGdefault@bf=\textbf\def\PYGdefault@tc##1{\textcolor[rgb]{0.00,0.50,0.00}{##1}}}
  \@namedef{PYGdefault@tok@nd}{\def\PYGdefault@tc##1{\textcolor[rgb]{0.67,0.13,1.00}{##1}}}
  \@namedef{PYGdefault@tok@s}{\def\PYGdefault@tc##1{\textcolor[rgb]{0.73,0.13,0.13}{##1}}}
  \@namedef{PYGdefault@tok@sd}{\let\PYGdefault@it=\textit\def\PYGdefault@tc##1{\textcolor[rgb]{0.73,0.13,0.13}{##1}}}
  \@namedef{PYGdefault@tok@si}{\let\PYGdefault@bf=\textbf\def\PYGdefault@tc##1{\textcolor[rgb]{0.73,0.40,0.53}{##1}}}
  \@namedef{PYGdefault@tok@se}{\let\PYGdefault@bf=\textbf\def\PYGdefault@tc##1{\textcolor[rgb]{0.73,0.40,0.13}{##1}}}
  \@namedef{PYGdefault@tok@sr}{\def\PYGdefault@tc##1{\textcolor[rgb]{0.73,0.40,0.53}{##1}}}
  \@namedef{PYGdefault@tok@ss}{\def\PYGdefault@tc##1{\textcolor[rgb]{0.10,0.09,0.49}{##1}}}
  \@namedef{PYGdefault@tok@sx}{\def\PYGdefault@tc##1{\textcolor[rgb]{0.00,0.50,0.00}{##1}}}
  \@namedef{PYGdefault@tok@m}{\def\PYGdefault@tc##1{\textcolor[rgb]{0.40,0.40,0.40}{##1}}}
  \@namedef{PYGdefault@tok@gh}{\let\PYGdefault@bf=\textbf\def\PYGdefault@tc##1{\textcolor[rgb]{0.00,0.00,0.50}{##1}}}
  \@namedef{PYGdefault@tok@gu}{\let\PYGdefault@bf=\textbf\def\PYGdefault@tc##1{\textcolor[rgb]{0.50,0.00,0.50}{##1}}}
  \@namedef{PYGdefault@tok@gd}{\def\PYGdefault@tc##1{\textcolor[rgb]{0.63,0.00,0.00}{##1}}}
  \@namedef{PYGdefault@tok@gi}{\def\PYGdefault@tc##1{\textcolor[rgb]{0.00,0.63,0.00}{##1}}}
  \@namedef{PYGdefault@tok@gr}{\def\PYGdefault@tc##1{\textcolor[rgb]{1.00,0.00,0.00}{##1}}}
  \@namedef{PYGdefault@tok@ge}{\let\PYGdefault@it=\textit}
  \@namedef{PYGdefault@tok@gs}{\let\PYGdefault@bf=\textbf}
  \@namedef{PYGdefault@tok@gp}{\let\PYGdefault@bf=\textbf\def\PYGdefault@tc##1{\textcolor[rgb]{0.00,0.00,0.50}{##1}}}
  \@namedef{PYGdefault@tok@go}{\def\PYGdefault@tc##1{\textcolor[rgb]{0.53,0.53,0.53}{##1}}}
  \@namedef{PYGdefault@tok@gt}{\def\PYGdefault@tc##1{\textcolor[rgb]{0.00,0.27,0.87}{##1}}}
  \@namedef{PYGdefault@tok@err}{\def\PYGdefault@bc##1{{\setlength{\fboxsep}{\string -\fboxrule}\fcolorbox[rgb]{1.00,0.00,0.00}{1,1,1}{\strut ##1}}}}
  \@namedef{PYGdefault@tok@kc}{\let\PYGdefault@bf=\textbf\def\PYGdefault@tc##1{\textcolor[rgb]{0.00,0.50,0.00}{##1}}}
  \@namedef{PYGdefault@tok@kd}{\let\PYGdefault@bf=\textbf\def\PYGdefault@tc##1{\textcolor[rgb]{0.00,0.50,0.00}{##1}}}
  \@namedef{PYGdefault@tok@kn}{\let\PYGdefault@bf=\textbf\def\PYGdefault@tc##1{\textcolor[rgb]{0.00,0.50,0.00}{##1}}}
  \@namedef{PYGdefault@tok@kr}{\let\PYGdefault@bf=\textbf\def\PYGdefault@tc##1{\textcolor[rgb]{0.00,0.50,0.00}{##1}}}
  \@namedef{PYGdefault@tok@bp}{\def\PYGdefault@tc##1{\textcolor[rgb]{0.00,0.50,0.00}{##1}}}
  \@namedef{PYGdefault@tok@fm}{\def\PYGdefault@tc##1{\textcolor[rgb]{0.00,0.00,1.00}{##1}}}
  \@namedef{PYGdefault@tok@vc}{\def\PYGdefault@tc##1{\textcolor[rgb]{0.10,0.09,0.49}{##1}}}
  \@namedef{PYGdefault@tok@vg}{\def\PYGdefault@tc##1{\textcolor[rgb]{0.10,0.09,0.49}{##1}}}
  \@namedef{PYGdefault@tok@vi}{\def\PYGdefault@tc##1{\textcolor[rgb]{0.10,0.09,0.49}{##1}}}
  \@namedef{PYGdefault@tok@vm}{\def\PYGdefault@tc##1{\textcolor[rgb]{0.10,0.09,0.49}{##1}}}
  \@namedef{PYGdefault@tok@sa}{\def\PYGdefault@tc##1{\textcolor[rgb]{0.73,0.13,0.13}{##1}}}
  \@namedef{PYGdefault@tok@sb}{\def\PYGdefault@tc##1{\textcolor[rgb]{0.73,0.13,0.13}{##1}}}
  \@namedef{PYGdefault@tok@sc}{\def\PYGdefault@tc##1{\textcolor[rgb]{0.73,0.13,0.13}{##1}}}
  \@namedef{PYGdefault@tok@dl}{\def\PYGdefault@tc##1{\textcolor[rgb]{0.73,0.13,0.13}{##1}}}
  \@namedef{PYGdefault@tok@s2}{\def\PYGdefault@tc##1{\textcolor[rgb]{0.73,0.13,0.13}{##1}}}
  \@namedef{PYGdefault@tok@sh}{\def\PYGdefault@tc##1{\textcolor[rgb]{0.73,0.13,0.13}{##1}}}
  \@namedef{PYGdefault@tok@s1}{\def\PYGdefault@tc##1{\textcolor[rgb]{0.73,0.13,0.13}{##1}}}
  \@namedef{PYGdefault@tok@mb}{\def\PYGdefault@tc##1{\textcolor[rgb]{0.40,0.40,0.40}{##1}}}
  \@namedef{PYGdefault@tok@mf}{\def\PYGdefault@tc##1{\textcolor[rgb]{0.40,0.40,0.40}{##1}}}
  \@namedef{PYGdefault@tok@mh}{\def\PYGdefault@tc##1{\textcolor[rgb]{0.40,0.40,0.40}{##1}}}
  \@namedef{PYGdefault@tok@mi}{\def\PYGdefault@tc##1{\textcolor[rgb]{0.40,0.40,0.40}{##1}}}
  \@namedef{PYGdefault@tok@il}{\def\PYGdefault@tc##1{\textcolor[rgb]{0.40,0.40,0.40}{##1}}}
  \@namedef{PYGdefault@tok@mo}{\def\PYGdefault@tc##1{\textcolor[rgb]{0.40,0.40,0.40}{##1}}}
  \@namedef{PYGdefault@tok@ch}{\let\PYGdefault@it=\textit\def\PYGdefault@tc##1{\textcolor[rgb]{0.25,0.50,0.50}{##1}}}
  \@namedef{PYGdefault@tok@cm}{\let\PYGdefault@it=\textit\def\PYGdefault@tc##1{\textcolor[rgb]{0.25,0.50,0.50}{##1}}}
  \@namedef{PYGdefault@tok@cpf}{\let\PYGdefault@it=\textit\def\PYGdefault@tc##1{\textcolor[rgb]{0.25,0.50,0.50}{##1}}}
  \@namedef{PYGdefault@tok@c1}{\let\PYGdefault@it=\textit\def\PYGdefault@tc##1{\textcolor[rgb]{0.25,0.50,0.50}{##1}}}
  \@namedef{PYGdefault@tok@cs}{\let\PYGdefault@it=\textit\def\PYGdefault@tc##1{\textcolor[rgb]{0.25,0.50,0.50}{##1}}}

  \makeatother

\makeatletter
  \def\PYG@reset{\let\PYG@it=\relax \let\PYG@bf=\relax%
      \let\PYG@ul=\relax \let\PYG@tc=\relax%
      \let\PYG@bc=\relax \let\PYG@ff=\relax}
  \def\PYG@tok#1{\csname PYG@tok@#1\endcsname}
  \def\PYG@toks#1+{\ifx\relax#1\empty\else%
      \PYG@tok{#1}\expandafter\PYG@toks\fi}
  \def\PYG@do#1{\PYG@bc{\PYG@tc{\PYG@ul{%
      \PYG@it{\PYG@bf{\PYG@ff{#1}}}}}}}
  \def\PYG#1#2{\PYG@reset\PYG@toks#1+\relax+\PYG@do{#2}}
  
  \@namedef{PYG@tok@w}{\def\PYG@tc##1{\textcolor[rgb]{0.73,0.73,0.73}{##1}}}
  \@namedef{PYG@tok@c}{\let\PYG@it=\textit\def\PYG@tc##1{\textcolor[rgb]{0.25,0.50,0.50}{##1}}}
  \@namedef{PYG@tok@cp}{\def\PYG@tc##1{\textcolor[rgb]{0.74,0.48,0.00}{##1}}}
  \@namedef{PYG@tok@k}{\let\PYG@bf=\textbf\def\PYG@tc##1{\textcolor[rgb]{0.00,0.50,0.00}{##1}}}
  \@namedef{PYG@tok@kp}{\def\PYG@tc##1{\textcolor[rgb]{0.00,0.50,0.00}{##1}}}
  \@namedef{PYG@tok@kt}{\def\PYG@tc##1{\textcolor[rgb]{0.69,0.00,0.25}{##1}}}
  \@namedef{PYG@tok@o}{\def\PYG@tc##1{\textcolor[rgb]{0.40,0.40,0.40}{##1}}}
  \@namedef{PYG@tok@ow}{\let\PYG@bf=\textbf\def\PYG@tc##1{\textcolor[rgb]{0.67,0.13,1.00}{##1}}}
  \@namedef{PYG@tok@nb}{\def\PYG@tc##1{\textcolor[rgb]{0.00,0.50,0.00}{##1}}}
  \@namedef{PYG@tok@nf}{\def\PYG@tc##1{\textcolor[rgb]{0.00,0.00,1.00}{##1}}}
  \@namedef{PYG@tok@nc}{\let\PYG@bf=\textbf\def\PYG@tc##1{\textcolor[rgb]{0.00,0.00,1.00}{##1}}}
  \@namedef{PYG@tok@nn}{\let\PYG@bf=\textbf\def\PYG@tc##1{\textcolor[rgb]{0.00,0.00,1.00}{##1}}}
  \@namedef{PYG@tok@ne}{\let\PYG@bf=\textbf\def\PYG@tc##1{\textcolor[rgb]{0.82,0.25,0.23}{##1}}}
  \@namedef{PYG@tok@nv}{\def\PYG@tc##1{\textcolor[rgb]{0.10,0.09,0.49}{##1}}}
  \@namedef{PYG@tok@no}{\def\PYG@tc##1{\textcolor[rgb]{0.53,0.00,0.00}{##1}}}
  \@namedef{PYG@tok@nl}{\def\PYG@tc##1{\textcolor[rgb]{0.63,0.63,0.00}{##1}}}
  \@namedef{PYG@tok@ni}{\let\PYG@bf=\textbf\def\PYG@tc##1{\textcolor[rgb]{0.60,0.60,0.60}{##1}}}
  \@namedef{PYG@tok@na}{\def\PYG@tc##1{\textcolor[rgb]{0.49,0.56,0.16}{##1}}}
  \@namedef{PYG@tok@nt}{\let\PYG@bf=\textbf\def\PYG@tc##1{\textcolor[rgb]{0.00,0.50,0.00}{##1}}}
  \@namedef{PYG@tok@nd}{\def\PYG@tc##1{\textcolor[rgb]{0.67,0.13,1.00}{##1}}}
  \@namedef{PYG@tok@s}{\def\PYG@tc##1{\textcolor[rgb]{0.73,0.13,0.13}{##1}}}
  \@namedef{PYG@tok@sd}{\let\PYG@it=\textit\def\PYG@tc##1{\textcolor[rgb]{0.73,0.13,0.13}{##1}}}
  \@namedef{PYG@tok@si}{\let\PYG@bf=\textbf\def\PYG@tc##1{\textcolor[rgb]{0.73,0.40,0.53}{##1}}}
  \@namedef{PYG@tok@se}{\let\PYG@bf=\textbf\def\PYG@tc##1{\textcolor[rgb]{0.73,0.40,0.13}{##1}}}
  \@namedef{PYG@tok@sr}{\def\PYG@tc##1{\textcolor[rgb]{0.73,0.40,0.53}{##1}}}
  \@namedef{PYG@tok@ss}{\def\PYG@tc##1{\textcolor[rgb]{0.10,0.09,0.49}{##1}}}
  \@namedef{PYG@tok@sx}{\def\PYG@tc##1{\textcolor[rgb]{0.00,0.50,0.00}{##1}}}
  \@namedef{PYG@tok@m}{\def\PYG@tc##1{\textcolor[rgb]{0.40,0.40,0.40}{##1}}}
  \@namedef{PYG@tok@gh}{\let\PYG@bf=\textbf\def\PYG@tc##1{\textcolor[rgb]{0.00,0.00,0.50}{##1}}}
  \@namedef{PYG@tok@gu}{\let\PYG@bf=\textbf\def\PYG@tc##1{\textcolor[rgb]{0.50,0.00,0.50}{##1}}}
  \@namedef{PYG@tok@gd}{\def\PYG@tc##1{\textcolor[rgb]{0.63,0.00,0.00}{##1}}}
  \@namedef{PYG@tok@gi}{\def\PYG@tc##1{\textcolor[rgb]{0.00,0.63,0.00}{##1}}}
  \@namedef{PYG@tok@gr}{\def\PYG@tc##1{\textcolor[rgb]{1.00,0.00,0.00}{##1}}}
  \@namedef{PYG@tok@ge}{\let\PYG@it=\textit}
  \@namedef{PYG@tok@gs}{\let\PYG@bf=\textbf}
  \@namedef{PYG@tok@gp}{\let\PYG@bf=\textbf\def\PYG@tc##1{\textcolor[rgb]{0.00,0.00,0.50}{##1}}}
  \@namedef{PYG@tok@go}{\def\PYG@tc##1{\textcolor[rgb]{0.53,0.53,0.53}{##1}}}
  \@namedef{PYG@tok@gt}{\def\PYG@tc##1{\textcolor[rgb]{0.00,0.27,0.87}{##1}}}
  \@namedef{PYG@tok@err}{\def\PYG@bc##1{{\setlength{\fboxsep}{\string -\fboxrule}\fcolorbox[rgb]{1.00,0.00,0.00}{1,1,1}{\strut ##1}}}}
  \@namedef{PYG@tok@kc}{\let\PYG@bf=\textbf\def\PYG@tc##1{\textcolor[rgb]{0.00,0.50,0.00}{##1}}}
  \@namedef{PYG@tok@kd}{\let\PYG@bf=\textbf\def\PYG@tc##1{\textcolor[rgb]{0.00,0.50,0.00}{##1}}}
  \@namedef{PYG@tok@kn}{\let\PYG@bf=\textbf\def\PYG@tc##1{\textcolor[rgb]{0.00,0.50,0.00}{##1}}}
  \@namedef{PYG@tok@kr}{\let\PYG@bf=\textbf\def\PYG@tc##1{\textcolor[rgb]{0.00,0.50,0.00}{##1}}}
  \@namedef{PYG@tok@bp}{\def\PYG@tc##1{\textcolor[rgb]{0.00,0.50,0.00}{##1}}}
  \@namedef{PYG@tok@fm}{\def\PYG@tc##1{\textcolor[rgb]{0.00,0.00,1.00}{##1}}}
  \@namedef{PYG@tok@vc}{\def\PYG@tc##1{\textcolor[rgb]{0.10,0.09,0.49}{##1}}}
  \@namedef{PYG@tok@vg}{\def\PYG@tc##1{\textcolor[rgb]{0.10,0.09,0.49}{##1}}}
  \@namedef{PYG@tok@vi}{\def\PYG@tc##1{\textcolor[rgb]{0.10,0.09,0.49}{##1}}}
  \@namedef{PYG@tok@vm}{\def\PYG@tc##1{\textcolor[rgb]{0.10,0.09,0.49}{##1}}}
  \@namedef{PYG@tok@sa}{\def\PYG@tc##1{\textcolor[rgb]{0.73,0.13,0.13}{##1}}}
  \@namedef{PYG@tok@sb}{\def\PYG@tc##1{\textcolor[rgb]{0.73,0.13,0.13}{##1}}}
  \@namedef{PYG@tok@sc}{\def\PYG@tc##1{\textcolor[rgb]{0.73,0.13,0.13}{##1}}}
  \@namedef{PYG@tok@dl}{\def\PYG@tc##1{\textcolor[rgb]{0.73,0.13,0.13}{##1}}}
  \@namedef{PYG@tok@s2}{\def\PYG@tc##1{\textcolor[rgb]{0.73,0.13,0.13}{##1}}}
  \@namedef{PYG@tok@sh}{\def\PYG@tc##1{\textcolor[rgb]{0.73,0.13,0.13}{##1}}}
  \@namedef{PYG@tok@s1}{\def\PYG@tc##1{\textcolor[rgb]{0.73,0.13,0.13}{##1}}}
  \@namedef{PYG@tok@mb}{\def\PYG@tc##1{\textcolor[rgb]{0.40,0.40,0.40}{##1}}}
  \@namedef{PYG@tok@mf}{\def\PYG@tc##1{\textcolor[rgb]{0.40,0.40,0.40}{##1}}}
  \@namedef{PYG@tok@mh}{\def\PYG@tc##1{\textcolor[rgb]{0.40,0.40,0.40}{##1}}}
  \@namedef{PYG@tok@mi}{\def\PYG@tc##1{\textcolor[rgb]{0.40,0.40,0.40}{##1}}}
  \@namedef{PYG@tok@il}{\def\PYG@tc##1{\textcolor[rgb]{0.40,0.40,0.40}{##1}}}
  \@namedef{PYG@tok@mo}{\def\PYG@tc##1{\textcolor[rgb]{0.40,0.40,0.40}{##1}}}
  \@namedef{PYG@tok@ch}{\let\PYG@it=\textit\def\PYG@tc##1{\textcolor[rgb]{0.25,0.50,0.50}{##1}}}
  \@namedef{PYG@tok@cm}{\let\PYG@it=\textit\def\PYG@tc##1{\textcolor[rgb]{0.25,0.50,0.50}{##1}}}
  \@namedef{PYG@tok@cpf}{\let\PYG@it=\textit\def\PYG@tc##1{\textcolor[rgb]{0.25,0.50,0.50}{##1}}}
  \@namedef{PYG@tok@c1}{\let\PYG@it=\textit\def\PYG@tc##1{\textcolor[rgb]{0.25,0.50,0.50}{##1}}}
  \@namedef{PYG@tok@cs}{\let\PYG@it=\textit\def\PYG@tc##1{\textcolor[rgb]{0.25,0.50,0.50}{##1}}}

  \def\PYGZus{\char`\_}

  \makeatother

\begin{document}


\title{\bf \SYSNAME: Virtualized Multi-Language Runtime for \\ High-Density Serverless Platforms}

\author{Serhii Ivanenko}
\orcid{0000-0002-4961-2679}
\email{serhii.ivanenko@tecnico.ulisboa.pt}
\affiliation{%
  \institution{INESC-ID, Instituto Superior Técnico, University of Lisbon}
  \city{Lisbon}
  \country{Portugal}
}

\author{Vasyl Lanko}
\orcid{0009-0002-6358-6742}
\email{vasyl.lanko@tecnico.ulisboa.pt}
\affiliation{%
  \institution{INESC-ID, Instituto Superior Técnico, University of Lisbon}
  \city{Lisbon}
  \country{Portugal}
}

\author{Rudi Horn}
\orcid{0000-0001-6164-2208}
\email{rudi.horn@oracle.com}
\affiliation{%
 \institution{Oracle Labs}
 \city{Zurich}
 \country{Switzerland}
}

\author{Vojin Jovanovic}
\orcid{0009-0002-4233-2401}
\email{vojin.jovanovic@oracle.com}
\affiliation{%
 \institution{Oracle Labs}
 \city{Zurich}
 \country{Switzerland}
}

\author{Rodrigo Bruno}
\orcid{0000-0003-1578-5149}
\email{rodrigo.bruno@tecnico.ulisboa.pt}
\affiliation{%
  \institution{INESC-ID, Instituto Superior Técnico, University of Lisbon}
  \city{Lisbon}
  \country{Portugal}
}
\pagestyle{plain}

\begin{abstract}
Serverless is an attractive computing model that offers seamless scalability and elasticity; it takes the infrastructure management burden away from users and enables a pay-as-you-use billing model. As a result, serverless is becoming increasingly popular to support highly elastic and bursty workloads. However, existing platforms are supported by bloated virtualization stacks, which, combined with bursty and irregular invocations, lead to high memory and latency overheads.
%

To reduce the virtualization stack bloat, we propose \SYSNAME, a virtualized multi-language runtime and platform capable of hosting multiple sandboxes running concurrently. To fully leverage \SYSNAME's virtualized runtime, we revisit the existing serverless platform design to make it colocation-aware across owners and functions, and to feature a caching layer of pre-allocated \SYSNAME instances that can be used by different functions written in different languages to reduce cold starts. We also propose a snapshotting mechanism to checkpoint and restore individual sandboxes.

By consolidating multiple serverless function invocations through \SYSNAME, we improve the overall function density (ops/GB-sec) by 2.41$\times$ on average compared to OpenWhisk runtimes, the state-of-the-art single-language runtimes used in most serverless platforms, and by 1.43$\times$ on average compared to Knative runtimes supporting invocation colocation within the same function. When reproducing the Azure Functions trace, our serverless platform operating \SYSNAME instances reduces the overall memory footprint by 21.3-43.9\% compared to operating OpenWhisk instances and by 14.5-30\% compared to operating Knative instances. \SYSNAME eliminates cold starts thanks to the pool of pre-warmed runtime instances, reducing p99 latency by 45.3-375.5$\times$ compared to OpenWhisk and by 1.9-51.4$\times$ compared to Knative.
\end{abstract}

\maketitle

\section{Introduction}
\label{sec:intro}

Serverless emerged as a new attractive computing model in which applications are composed of lightweight and fast-executing snippets of code commonly referred to as functions. Serverless platforms offer great scalability and elasticity, taking the infrastructure management burden away from users, and enabling a \textit{pay-as-you-use} billing model where users pay only for the time the function is running~\cite{catro:2019,smith:2021}. As a result, serverless is becoming increasingly popular to deploy elastic and bursty workloads such as video and image processing~\cite{fouladi:2017}, Machine Learning~\cite{carreira:2019}, linear algebra~\cite{shankar:2020}, data analytics~\cite{perron:2020,mueller:2020}, and web-apps/microservices~\cite{gan:2019}. However, as the utilization of serverless grows, platforms are faced with high memory overheads and long tail latencies. We attribute these to a combination of two factors:

\begin{figure}[t]
  \centering
  \includegraphics[width=\linewidth]{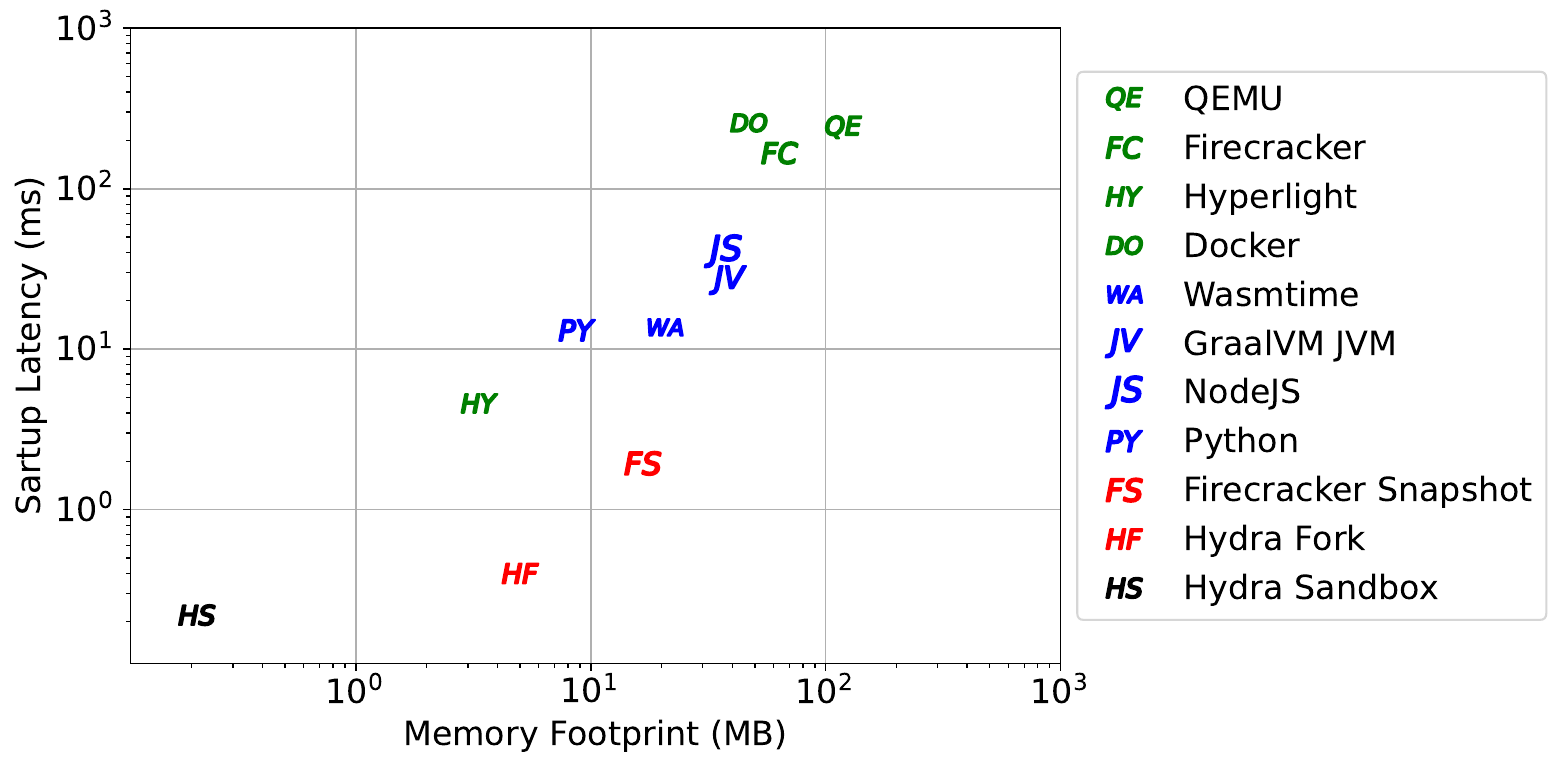}
    \caption{Performance landscape of different virtualization systems (\textcolor{teal}{\textbf{green}}), language runtimes (\textcolor{blue}{\textbf{blue}}), impact of forking and snapshotting (\textcolor{red}{\textbf{red}}), and \SYSNAME Sandbox (\textbf{black}).
  }
  \label{fig:virtualization-performance}
\end{figure}

\textbf{1. Serverless runs on bloated virtualization stacks.} A container (e.g., Docker~\cite{docker}) or a Virtual Machine (e.g., QEMU~\cite{qemu}, Firecracker~\cite{agache:2020}), and a language runtime (e.g., Node.js, CPython, or JVM) running on top of it are the most commonly used virtualization stacks. For each invocation handled by the platform, a separate virtualization stack is provisioned. State-of-the-art container/VM implementations lead to a long startup time and a high memory footprint inflicting a non-negligible overhead on lightweight and fast-execution serverless functions. The same applies to language runtimes such as CPython, JVM, and Node.js, the three most popular native language implementations that, according to a recent study~\cite{newrelic}, account for approximately 90\% of the total number of function invocations in Amazon Lambda~\cite{lambda}. These runtimes were developed to host long-running applications and, therefore, employ effective but time and resource-consuming techniques to optimize for long-term performance~\cite{carreira:2021,kohli:2024}. The combined memory footprint and startup latency of a container/VM and a runtime can easily dominate the total memory footprint and startup latency for serverless functions, that have been consistently reported as lightweight and fast executing in public traces~\cite{sharad:2020,joosen:2023,wang:2021}.

\begin{figure}[t]
  \centering
  \includegraphics[width=\linewidth]{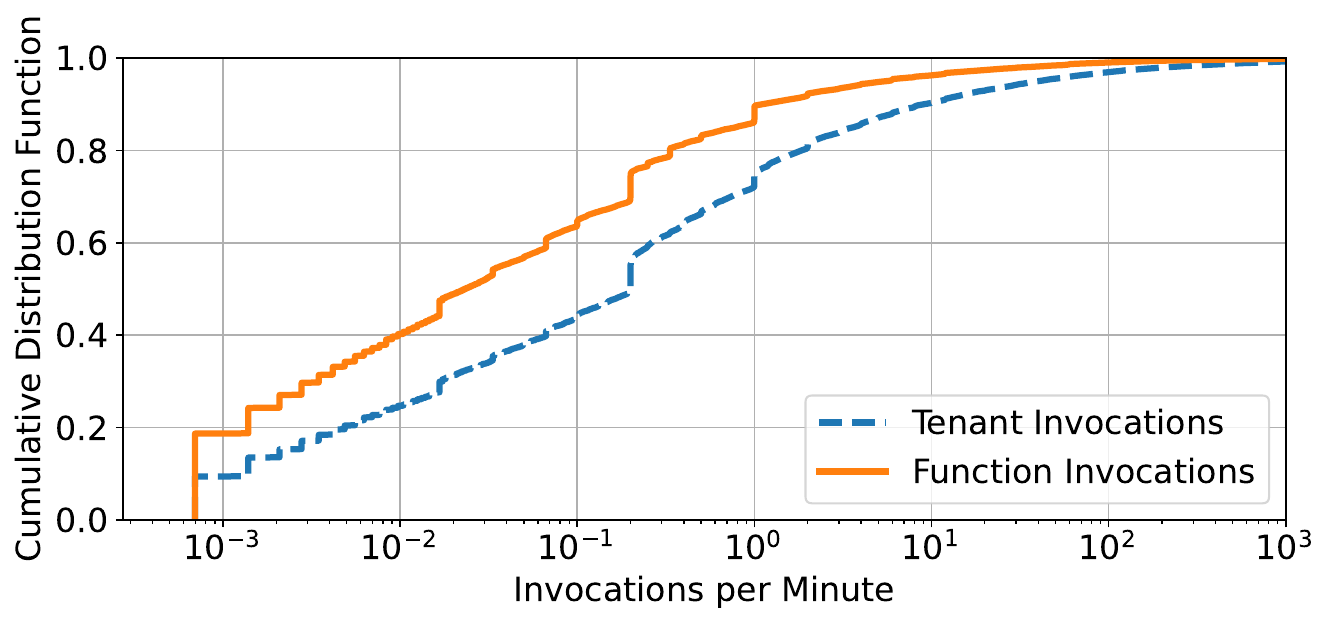}
  \caption{CDF of the average function invocations per minute during a full day (day 2 from Azure traces~\cite{sharad:2020}). Tenant invocations (aggregate invocations of all functions from the same tenant) show that most tenants deploy multiple functions.}
  \label{fig:azure:cdf}
\end{figure}

\begin{figure}[t]
  \centering
  \includegraphics[width=\linewidth]{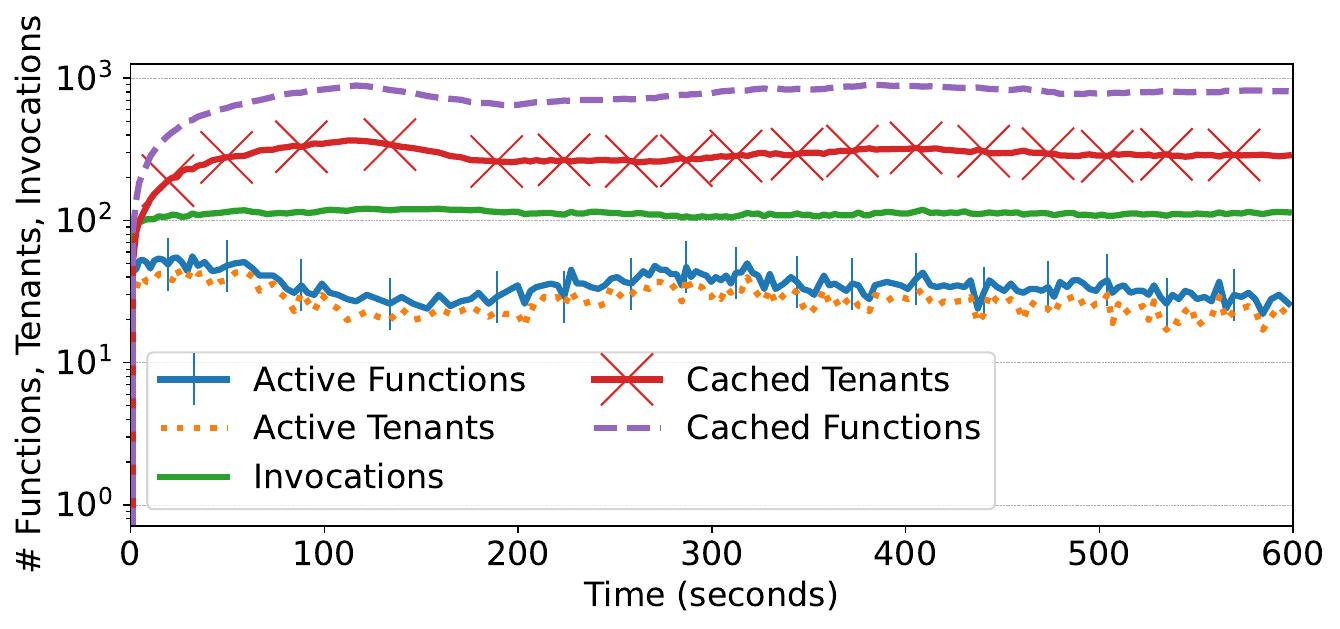}
  \caption{Invocations on a 10-minute segment (from Azure traces~\cite{sharad:2020}) scaled down to accommodate 16 GB of memory used by invocations. The plot also includes the number of active functions (number of functions being invoked) and tenants (aggregate number of active functions from the same tenant). The effects of caching are also depicted by including the number of functions (and tenants) that would be cached. 
  }
  \label{fig:azure:usersfunctions}
\end{figure}

\textbf{2. Serverless workloads are bursty.} Serverless workloads include periods of very low activity with only a few invocations, followed by periods of higher activity with multiple function invocations~\cite{sharad:2020}. Figure~\ref{fig:azure:cdf} shows how sparse the invocations of each function are; only 11\% of the functions will receive on average at least 1 invocation per minute, meaning that most virtualization stacks are highly underutilized. As a result of these factors, platforms resort to a trade-off between high memory overheads to cache virtualization stacks (waiting for the next invocation) and long tail invocation latencies inflicted by the cost of launching a new virtualization stack (cold start). Figure~\ref{fig:azure:usersfunctions} depicts the effect of caching functions recently invoked, a commonly used technique in both production systems such as Amazon Lambda, Google Cloud Functions, and Azure Functions, and open source serverless platforms such as OpenWhisk and Knative~\cite{wang:2018,lin:2025}. The results demonstrate that, as a consequence of the sparsity of the invocations of each function, the number of cached functions is much higher than the number of invocations. In turn, platforms need to carefully balance the size of the cache (memory overhead) with the number of function cold starts (tail latency).

Checkpoint/Restore~\cite{shin:2022,ustiugov:2021,silva:2020} (C/R) and runtime forking~\cite{akkus:2018,oakes:2018} have been extensively studied as a way to reduce memory footprint and tail latency. However, while both C/R and runtime forking reduce memory and startup latency by sharing a common stack (see Figure~\ref{fig:virtualization-performance}), both techniques optimize invocations of a single function, requiring separate virtualization stacks when multiple functions are invoked. Other techniques already virtualize language runtime~\cite{dukic:2020} but only allow single function invocations to co-execute on the same runtime, therefore requiring separate virtualization stacks for different functions.

In this paper, we propose \SYSNAME, a runtime and platform capable of sharing a single virtualization stack across concurrent invocations of multiple functions, even if written in different languages. \SYSNAME's runtime supports both AOT-compiled and JIT-compiled functions, and function invocations are executed in lightweight sandboxes that can be checkpointed and restored individually. In addition, we propose revisiting the serverless platform design to make it colocation-aware, allowing it to be configured with different colocation policies such as colocating all functions from the same tenant, or only colocating invocations for the same function. 


On a number of established serverless function benchmarks~\cite{dukic:2020,copik:2021}, we demonstrate how \SYSNAME's lightweight sandboxes increase function density (ops/GB-sec) in single-function scenarios compared to process forking, running functions in separate stacks, and state-of-the-art serverless runtimes. To study the impact of \SYSNAME on real-world platforms, we use \SYSNAME to serve the Azure Function trace~\cite{sharad:2020}. Overall, \SYSNAME reduces memory consumption by 21.3-43.9\% compared to operating OpenWhisk instances and by 14.5-30\% compared to operating Knative instances; it completely eliminates cold starts, and reduces p99 latency by 45.3-375.5$\times$ compared to OpenWhisk and by 1.9-51.4$\times$ compared to Knative when reproducing a public serverless trace~\cite{sharad:2020}.



In summary, this work offers the following contributions:
\begin{itemize}[leftmargin=*]

    \item we demonstrate that \textit{density} and \textit{elasticity} can be significantly improved with \SYSNAME. Density is improved by sharing additional components of the virtualization stack (\cref{sec:graalvisor}). Elasticity is improved restoring individual sandboxes from individual sandbox snapshots;

    \item we rethink the serverless platform design to incorporate colocation-aware scheduling and elastic runtime pools that effectively mitigate cold starts;
    
    \item we measure the elasticity and density benefits of \SYSNAME both at a local scale by comparing it to state-of-the-art language runtimes, and at a distributed scale by replaying the Azure Functions trace~\cite{sharad:2020} on a realistic platform (\cref{sec:evaluation}).
\end{itemize}

\section{Background and Motivation} \label{sec:background}
Serverless platforms inherit much of the virtualization technology that was designed to support long-running monolithic applications and microservices. In addition, serverless platforms have also traditionally been designed to work with virtualization stacks that do not support colocation. We argue that this infrastructure is not a good fit for lightweight, short-running, and bursty serverless functions.





\subsection{Traditional Virtualization Stacks}
\label{subsec:background:virtualization}

Existing virtualization stacks can be divided into two main groups: system-level virtualization, Virtual Machines (such as Xen~\cite{barham:2003}, QEMU~\cite{qemu}, and Firecracker~\cite{agache:2020}), and Operating System-level virtualization, containers (such as Docker~\cite{docker} and gVisor~\cite{gvisor}). VMs and containers virtualize the hardware and the Operating System, respectively. These technologies were designed to host long-running applications with high memory footprint demands that do not match most serverless functions~\cite{sharad:2020}. As a consequence, the slow start and high memory demands of such virtualization stacks easily dominate the function startup time and memory consumption.

There have been multiple attempts to fit traditional virtualization systems into serverless. \textbf{Runtime reuse} is a popular strategy to reduce the number of runtime startups by keeping a serverless worker (virtualization stack composed of a VM/container and a runtime) alive after a function invocation finishes~\cite{aws:warmstart,azure:warmstart,wang:2018}. If another invocation is received within a timeout limit, the worker is reused and no startup/initialization is required. A warm start refers to an invocation that is served by an already existing worker as opposed to a cold start which requires starting a new worker. Runtime reuse has, however, several limitations. First, it does not reduce resource redundancy as multiple concurrent requests still need to be served by independent workers. Second, only a small portion of serverless functions will benefit from runtime reuse as the vast majority of functions have very sparse invocations (see Figure~\ref{fig:azure:cdf}). Finally, it creates additional memory pressure to keep workers in memory after an invocation finishes (see Figure~\ref{fig:azure:usersfunctions}). 

\textbf{Snapshotting} and \textbf{forking} have also received much attention as a promising technique to reduce the memory footprint and startup latency of serverless functions. Restoring from a previous snapshot significantly reduces the startup time and memory footprint compared to a cold start~\cite{du:2020,ustiugov:2021,cadden:2020} but still requires a separate virtualization stack for each concurrent invocation and imposes storage/network overhead to manage snapshots~\cite{ao:2022}. Similarly, runtime forking~\cite{oakes:2018,akkus:2018} also reduces both memory footprint via copy-on-write and startup time by simply forking instead of launching a completely new virtualization stack. However, state-of-the-art runtimes such as CPython, Node.js, and JVM do not support forking out-of-the-box. For example, Garbage Collector and Just-In-Time compiler threads do not survive forking and require careful revival in the child process. 

\subsection{Lightweight Execution Environment}
\label{subsec:exec:environments}
Running multiple function invocations in a single runtime reduces startup latency and memory footprint of serverless functions by using a single virtualization stack. However, it requires enforcing resource isolation. CPU, file system, and network isolation can be guaranteed using Operating System-level primitives such as Linux Namespaces and Control Groups~\cite{cgroups}. Memory isolation requires function sandboxing. 
%
%
%

\begin{figure}[t]
  \centering
  \includegraphics[width=.9\linewidth]{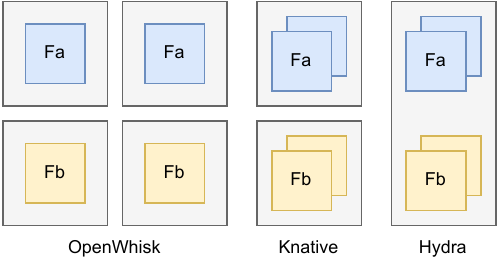}
  \caption{Virtualization stacks (gray boxes) needed to serve two concurrent invocations of two functions in different serverless platforms.}
  \label{fig:virt:arch}
\end{figure}

There are two main widely available sandboxing techniques in modern language engines: WebAssembly sandboxes and Memory Isolates. \textbf{WebAssembly}~\cite{haas:2017} is a binary-code format with memory safety guarantees that resulted from the evolution of previous Software Fault Isolation techniques such as NaCL~\cite{nacl}. These guarantees are enforced by restricting memory access to a single linear byte array with efficient bound checks during compilation and run-time. Runtimes supporting the execution of WebAssembly sandboxes can run multiple functions in a memory-isolated environment~\cite{shillaker:2020,gadepalli:2020}. However, modern WebAssembly runtimes do not include high-level languague support and thus, the function code would have to include the language engine. For example, existing support to run Python/JavaScript code requires functions to include CPython/V8 (or other engine) compiled into WebAssembly~\cite{shillaker:2020,fermyon}. This mechanism re-introduces redundancy as sandboxes do not share language engines, making it similar to OS-level virtualization stacks in Figure~\ref{fig:virt:arch}.

Language runtimes such as V8~\cite{v8} and GraalVM Native Image~\cite{graalvm} offer \textbf{Memory Isolates}, a memory segment that can be used to host the execution of a function. Similarly to WebAssembly sandboxes, an isolate is a linear memory segment used to keep application objects. References are resolved within the linear memory segment and the sandbox limits are enforced by the compiler and runtime. Isolates are, however, restricted to executing functions written in the language supported by the runtime, JavaScript for V8, and Java bytecode for GraalVM Native Image.

Platforms such as Photons~\cite{dukic:2020} and Knative~\cite{knative} propose weakening isolation to run concurrent requests in a single runtime with minimal or even no isolation at all. Photons offers transparent isolation for Java applications by applying Java bytecode instrumentation to automatically synchronize access to global shared state. Knative, on the other hand, leaves this concern to the developers, forcing serverless application developers to deal with concurrency and synchronization challenges.

\subsection{Runtime Impact on Serverless Platforms}
\label{subsec:background:impact}

The execution environment used to run functions can affect the infrastructure. For example, as shown in Figure~\ref{fig:virt:arch}, in order to run two concurrent invocations to two different functions, the serverless platform has to create four OpenWhisk~\cite{openwhisk} workers since OpenWhisk runtimes isolate each concurrent invocation in a single worker. If the platform uses Knative workers, it needs only two workers since Knative is capable of colocating concurrent invocations to the same function, providing no isolation within the same worker. \SYSNAME, on the other hand, requires a single worker to execute all invocations of both functions. 

Below, we illustrate how the type of runtime can affect the serverless design of the serverless platform:

\textbf{1. Limiting function resources.} Serverless workers running different functions share the same hardware. Therefore, it is crucial to limit worker resources, such as memory, to ensure a fair distribution of hardware between functions. If a serverless worker runs one request at a time, then its memory can be restricted to a fixed value. However, if the worker supports handling concurrent invocations  (as in Photons or Knative), its memory limits have to grow and shrink to accommodate function invocations bursts. However, modern language runtimes such as the JVM, and NodeJS do not support dynamic memory heap resizing, leading to significant memory waste~\cite{bruno:2018}.

\textbf{2. Pre-creating serverless workers.} Platforms like OpenWhisk feature function-generic language-specific container images for workers. A function is registered in the available worker before handling invocation. This enables the platform to pre-create a language-partitioned pool of workers to avoid creating a VM/container on the critical path of a cold start. However, this requires the platform to resize the worker's memory to the function's needs after creation. Language runtimes such as JVM probe the total available memory during startup to determine their heap size parameters, and reconfiguring the heap size at run-time is not trivial~\cite{bruno:2018}, preventing platforms from effectively resizing the worker. Other platforms, such as Knative, require developers to provide function-specific container images to start workers. This approach prevents platforms from keeping a pool of pre-created generic workers.



\subsection{Multi-Language Support}
Executing functions developed in different languages on top of a single runtime instance is also a key ingredient to reducing virtualization stack redundancy. WebAssembly accomplishes this by allowing multiple languages to target its binary code format or by including interpreters compiled into WebAssembly in the function code. On the other hand, memory isolates alone cannot offer multi-language capabilities on itself. An alternative approach is enabled by using Truffle~\cite{wuerthinger:2013}, a language implementation framework that allows developers to quickly write interpreters for a specific language. Truffle interpreters are then automatically processed and optimized, taking full advantage of Truffle's underlying JIT compiler. Truffle is supported by recent advances in language runtime and compiler literature~\cite{wimmer:2019,grimmer:2015a,latifi:2021,grimmer:2015b,zhang:2014,grimmer:2014,wuerthinger:2013,wimmer:2012,wuerthinger:2017,wuerthinger:2013,larisch:2018} and is designed to efficiently execute dynamic languages such as Python, JavaScript, and Ruby but also has support for LLVM bitcode~\cite{rigger:2016} and even WebAssembly~\cite{salim:2020}. Truffle allows functions written in different languages to execute in a Java runtime. However, Truffle code execution greatly suffers from long initialization times and high memory overhead (\cref{sec:eval:snapshot}), which easily nullifies any benefit obtained by virtualizing the runtime.

\section{\SYSNAME Language Runtime} \label{sec:graalvisor}

\begin{figure}[t]
  \centering
  \includegraphics[width=.9\linewidth]{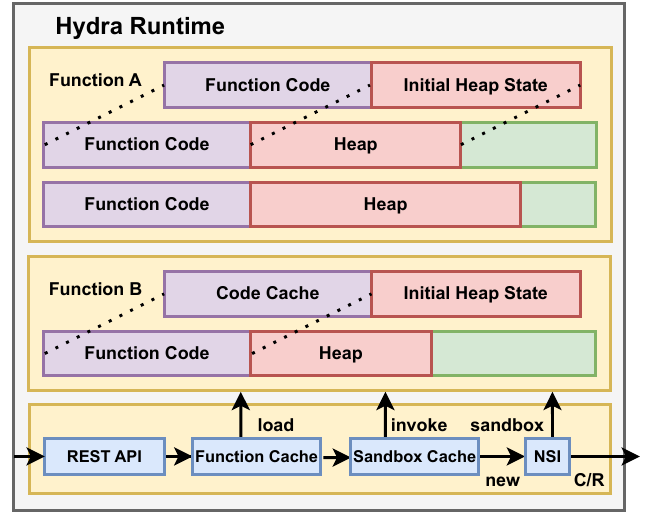}
  \caption{\SYSNAME running three function sandboxes.}
  \label{fig:graalvisor-arch}
\end{figure}

\SYSNAME is a language runtime capable of running multiple functions concurrently and in isolation, by confining each function into a lightweight sandbox. Figure~\ref{fig:graalvisor-arch} depicts an example of a \SYSNAME instance with 3 active sandboxes executing 2 different functions. Each sandbox runs invocations of a single function, one invocation at a time. At startup, \SYSNAME opens a network socket to serve REST requests (\cref{sec:graalvisor:interface}). Upon receiving an invocation request, \SYSNAME loads the function into memory if not already loaded, creates a new sandbox if no warm sandbox is available, and invokes the function inside the sandbox. Sandboxes of the same function share the same code cache and initial heap state (\cref{sec:graalvisor:isolates}). \SYSNAME offers multi-language support through Truffle (\cref{sec:graalvisor:polyglot}), and supports launching new sandboxes or restoring one from a previous snapshot (\cref{sec:graalvisor:cr}).

\subsection{Function Interface} \label{sec:graalvisor:interface}

\SYSNAME exposes a minimal canonical interface to the outside through a single end-point. Only three methods are required, function registration, invocation, and deregistation. This function interface is heavily inspired by existing function interfaces available in both commercial (Amazon Lambda~\cite{lambda}, Azure Functions~\cite{functions}, and Google Cloud Functions~\cite{gcf}), and open-source serverless platforms (such as OpenFaaS~\cite{openfaas}, OpenWhisk~\cite{openwhisk}, Knative~\cite{knative}). We expect that \SYSNAME requires minor to no function source code modifications to run existing serverless functions in \SYSNAME. 

\subsection{Function Sandboxes} \label{sec:graalvisor:isolates}
\SYSNAME sandboxes are based on GraalVM's Native Image~\cite{wimmer:2019} memory isolates which we extend to support on-demand sandbox creation, function invocation, and checkpoint/restore (\cref{sec:graalvisor:cr}). Native Image is an Ahead-Of-Time Java compiler that we use to compile user functions into shared libraries. \SYSNAME's design is independent from the sandbox implementation and could have been based on WebAssembly sandboxes, for example. 

Upon receiving a request to invoke a function, \SYSNAME loads the function shared library using \texttt{dlopen}, if not already loaded. The shared library loads into memory the function code cache and the initial heap state (see Figure~\ref{fig:graalvisor-arch}). The code cache contains the compiled user code that will be invoked through the registered entrypoint and the initial heap state contains a set of global values which are copied into the sandboxe's memory heap upon modification~\cite{wimmer:2019} (copy-on-write). Both the code cache and the initial heap state are shared among all sandboxes of a single function. Different functions have separate code caches and initial heap states.

Listing~\ref{lst:isolateapi} presents the ABI that \SYSNAME uses to create and invoke function sandboxes.~\footnote{\SYSNAME's ABI is built on top of Native Image shared library ABI.} The \texttt{create\_sandbox} creates a sandbox and returns a pointer to the new sandbox (through the second function argument). Once a sandbox is created, threads can be attached to it through the \texttt{attach\_thread} function. This function creates a new thread stack in the given sandbox and returns a pointer to the thread data structure (which can be later used to detach the thread from the sandbox by calling \texttt{detach\_thread}). Note that threads can be created in any sandbox and will start executing in the same sandbox as the parent thread.

\begin{listing}[!ht]
\begin{Verbatim}[commandchars=\\\{\}]
  \PYG{k+kt}{int} \PYG{n+nf}{create\PYGZus{}sandbox}\PYG{p}{(}\PYG{n}{sandbox\PYGZus{}params\PYGZus{}t}\PYG{o}{*}\PYG{p}{,} \PYG{n}{sandbox\PYGZus{}t}\PYG{o}{**}\PYG{p}{);}
  \PYG{k+kt}{int} \PYG{n+nf}{attach\PYGZus{}thread}\PYG{p}{(}\PYG{n}{sandbox\PYGZus{}t}\PYG{o}{*}\PYG{p}{,} \PYG{n}{sandbox\PYGZus{}thread\PYGZus{}t}\PYG{o}{**}\PYG{p}{);}
  \PYG{k+kt}{int} \PYG{n+nf}{invoke\PYGZus{}sandbox}\PYG{p}{(}\PYG{n}{sandbox\PYGZus{}thread\PYGZus{}t}\PYG{o}{*}\PYG{p}{,} \PYG{k+kt}{char}\PYG{o}{*}\PYG{p}{,} \PYG{k+kt}{char}\PYG{o}{*}\PYG{p}{,} \PYG{k+kt}{int}\PYG{p}{);}
  \PYG{k+kt}{int} \PYG{n+nf}{detach\PYGZus{}thread}\PYG{p}{(}\PYG{n}{sandbox\PYGZus{}thread\PYGZus{}t}\PYG{o}{*}\PYG{p}{);}
  \PYG{k+kt}{int} \PYG{n+nf}{tear\PYGZus{}down\PYGZus{}isolate}\PYG{p}{(}\PYG{n}{sandbox\PYGZus{}t}\PYG{o}{*}\PYG{p}{);}
  \end{Verbatim}


\caption{\SYSNAME Sandbox ABI in C.}
\label{lst:isolateapi}
\end{listing}

Functions are invoked through \texttt{invoke\_sandbox}, passing the target sandboxed thread handle where the request will be handled, a buffer where the arguments should be read from (for example, a JSON represented as a C string), and another buffer where result should be placed (also as a C string). The final argument indicates the maximum output size. All functions return an error code.

\subsection{Multi-Language Support in \SYSNAME}
\label{sec:graalvisor:polyglot}
Native Image only supports Java functions. To run functions written in languages other than Java, \SYSNAME takes advantage of Truffle~\cite{wuerthinger:2017}, a Java-based language implementation framework. We extend the Truffle framework to i) transparently load and configure the correct Truffle language, ii) load the function code into the Truffle engine, and iii) integrate it with \SYSNAME's sandbox ABI (introduced in Listing~\ref{lst:isolateapi}). Using Native Image, we AOT compile the Truffle language implementation bundled together with the function code into a dynamic library that is then registered into \SYSNAME. The resulting library supports \SYSNAME's sandbox ABI. The function entry point jumps to a call gate injected by us that enforces the correct translation of Java types into the target language types (used to pass invocation arguments and return values).

Truffle language implementations are interpreted and Just-In-Time (JIT) compiled. As a consequence, the code cache will be updated by the Truffle JIT compile that comes built-in the shared library. Multiple sandboxes running same function share the code cache this reusing already profiled and optimized code.




\subsection{\SYSNAME Sandbox Snapshots}
\label{sec:graalvisor:cr}
Function sandboxes require significant warmup time and memory, especially for JIT-compiled languages such as JavaScript. To reduce the startup time and memory footprint of these sandboxes, \SYSNAME is capable checkpointing function sandboxes. A snapshot includes the entire sandbox state such that all the code interpretation and JIT compilation overheads can be bypassed upon restore.

Checkpointing a single sandbox is not trivial as it requires saving all memory regions and resources that pertain to a single sandbox. Taking CRIU~\cite{criu} as a reference, C/R is done at the process level by inspecting the entire set of resources utilized by a process and dumping them into one or more files to be later used to restore the entire process. In \SYSNAME, this is not possible as we need to checkpoint a single sandbox and not the entire process. For instance, it is not possible to determine which memory mappings and file descriptors belong to a particular sandbox as the OS organizes resources by process. 

To overcome this challenge, \SYSNAME tracks the execution of a sandbox by intercepting system calls and tracking their side effects. By keeping track of all the resources utilized by a sandbox, it is possible to checkpoint its state and restore it later. \SYSNAME's sandbox checkpoint/restore design is focused on four main elements we found to be crucial for Java, Python, and JavaScript applications: memory mappings (used by language engines to allocate managed heaps), file descriptors, libc memory allocator (essential to many libc functions used by language engines), and threads running in the sandbox.

\subsubsection{Function Sandbox Checkpoint}
\SYSNAME utilizes \texttt{seccomp-bpf} to selectively intercept system calls by attaching a \texttt{seccomp-bpf}~\cite{seccomp} filter to the thread handling the function invocation. After installing the filter, the thread loads the function library with \texttt{dlopen}, creates a sandbox, and invokes the function code. Another thread is used track system call notifications and monitors system calls related to memory (such as \texttt{mmap}, \texttt{munmap}, and \texttt{mprotect}), files (such as \texttt{openat}, \texttt{dup}), and threads (\texttt{clone}, \texttt{clone3}, \texttt{exit}). We will focus on these three main classes of system calls and omit, for simplicity, other less important system calls and system calls variants. 

Upon receiving a system call notification, the monitor thread registers the arguments and the return value. This information is kept in a data structure in memory. In addition to keeping track of the received system calls, the monitor thread also keeps the current state of memory mappings, open file descriptors, and running threads belonging to the target sandbox.

\textbf{Memory mappings and file descriptors.} For memory-related system calls, the monitor keeps a list of memory mappings including not only the range but also their current permission and if the memory in that mapping was ever writable. \SYSNAME keeps this information at the page granularity. A similar table is kept for file-related operations, tracking which file descriptors are in use.

The monitor thread waits until the function invocation finishes and detaches the invoking thread from the sandbox to proceed with the checkpoint operation. At this point, the monitor starts iterating over the list of system calls to remove unnecessary entries. For example, a file may be opened, its contents read into memory, and then closed. In this case, \SYSNAME discards these \texttt{open} and \texttt{close} system calls as it targets restoring the final state sandbox. The same principle is applied to memory operations where \texttt{mmap} system calls whose return address was already \texttt{unmmap}ed can be safely ignored. The final list of system calls (including arguments and return values) is then written to a file used during the restore operation. For the benchmarks used in~\cref{sec:evaluation}, the final number of system calls was reduced by 47.1\%. on average.

The second checkpoint step is to iterate over all memory mappings and copy the memory contents to a file. We optimize this stage by copying only the mappings that could have been modified (i.e., write permission was enabled at some point in time). Note that all read-only memory will be restored when the system calls that loaded that memory are replayed (for instance, an \texttt{mmap} that loads a binary into memory). Two files result from a checkpoint operation: i) a file containing the set of system calls that need to be replayed upon a restore, and ii) the memory contents that need to be restored.

\textbf{Threads.} Similarly to memory and file descriptors, \SYSNAME keeps track of thread creation and destruction by monitoring system calls. However, saving the state of a thread requires stopping it and saving its CPU state. In \SYSNAME, we do this by freezing all threads  associated with the sandbox being checkpointed (this association is kept by monitoring system calls) and sending them a signal (we use \texttt{SIGUSR1} but other signals could be used). Background threads running in the sandbox are frozen while the checkpoint operation is in progress, and are resumed after, allowing the sandbox to receive further invocations.

After receiving a signal, and once inside the signal handler installed by \SYSNAME, \SYSNAME saves the \texttt{ucontext\_t} passed as argument into the handler. This data structure contains the CPU state of the thread (general-purpose register values, stack pointer, instruction pointer, etc.) necessary to restore the thread later on. Note that the thread state is saved while the thread is inside the signal handler as this will be important for the restore operation.

\textbf{Memory allocator.} Besides memory mappings resulting from \texttt{mmap}, \texttt{libc}'s memory allocator, \texttt{malloc}, may also be called to allocate memory. For instance, creating a thread with \texttt{pthread\_create} will allocate memory through \texttt{malloc}. Since multiple sandboxes share the same underlying \texttt{libc}, it becomes important to separate the memory chunks allocated with \texttt{libc}'s allocator for different sandboxes.

To separate the allocated memory for different sandboxes, \SYSNAME offers a virtualized memory allocator that overrides \texttt{libc}'s memory allocation interface (\texttt{malloc}, \texttt{free}, etc.). \SYSNAME's memory allocator is based on the Doug Lea's memory allocator~\cite{malloc}, which includes support for memory spaces. \SYSNAME's memory allocator uses one memory space per sandbox and supports checkpoint and restore operations. 
Upon a call to the allocator (e.g., a call to \texttt{malloc}) the allocator starts by checking if the thread already belongs to a memory space. This check is based on a thread-local variable (fast-path), or by checking a table kept in the allocator metadata (slow-path). When a checkpoint operation is requested, the allocator writes to disk all of its internal metadata regarding a particular sandbox.

\subsubsection{Function Sandbox Restore}
\SYSNAME can restore a function sandbox (including the code cache and the initial heap state) from the snapshot produced during a checkpoint operation. \SYSNAME starts by loading the file that contains the system calls that need to be replayed. For each system call, the restoring thread replays the system call and confirms that the output is the one expected (e.g., checking that the file descriptor number is the same as the one obtained in the original execution before the checkpoint). Note that for \texttt{mmap} system calls, \SYSNAME forces the address to be the same as the one returned by the original \texttt{mmap} system call by passing the target address and the \texttt{MAP\_FIXED} flag. After replaying all memory map and file-related system calls, memory contents and open files match the state of the original sandbox.

\begin{figure}[t]
  \centering
  \includegraphics[width=\linewidth]{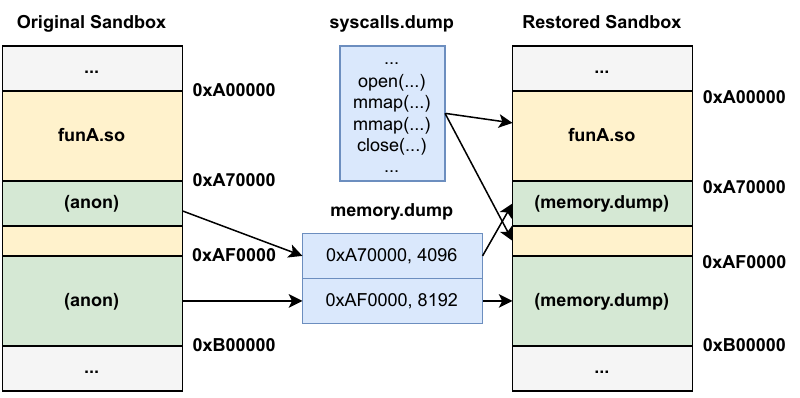}
  \caption{Function checkpoint/restore example. Non-writable memory is not included in the memory snapshot file that includes the address and size of each mapping to restore. The system call snapshot file includes all recorded system calls including arguments and return values.}
  \label{fig:svm-snapshto}
\end{figure}

After this initial step that restores file descriptors, file-based memory mappings, and read-only memory mappings, the restoring thread loads the \texttt{memory.dump} file contents to the expected memory positions (see Figure~\ref{fig:svm-snapshto}). This file includes only memory mappings recorded in the original execution that may have been dirtied. This technique significantly speeds up the restore operation at the expense of some additional page faults during the first invocation of the function (note that this could be further optimized by prefetching the pages estimated to be necessary as done in previous work~\cite{ustiugov:2021}).

In the third step, \SYSNAME restores the memory allocator state by loading the memory space metadata into \SYSNAME's memory allocator. And finally, \SYSNAME restores sandbox threads. Thread restoration is achieved with the help of the \texttt{sigreturn} system call that receives the \texttt{ucontext\_t} structure passed into the signal handler in the checkpoint step. The \texttt{sigreturn} system call allows the kernel to restore the state of the thread by setting all its registers. The thread is restored in its original application as if it had just received a signal. System calls interrupted during the checkpoint step (e.g. \texttt{futex}) are automatically re-started upon a restore. Finally, we ensure that threads are restored using the same thread id by controlling the value in \texttt{ns\_last\_pid} file in the \texttt{proc} file system.


\subsubsection{Restoring Multiple Snapshots}
The C/R mechanism described so far can easily fail if two function snapshots require the same virtual address range. This can happen, for example, if two calls to \texttt{mmap} return the same address in two different executions of \SYSNAME where each function was individually checkpointed. Without a reliable mechanism to address this issue, restore operations can easily fail if the virtual memory is already being used.

To avoid this scenario, \SYSNAME controls which memory ranges are assigned to each sandbox. This is possible by intercepting and adjusting the \texttt{mmap} arguments and forcing each function to operate in disjoint memory ranges. Note that this mechanism works as most language engines reserve large blocks of memory which are then used internally to feed allocation pools (e.g., memory heaps). \SYSNAME's memory allocator also relies on \texttt{mmap} to serve memory arenas.

\SYSNAME reserves 32 GBs of virtual memory for each function, although this value is configurable. 
We use a unique seed value that is used to pick a 32 GB range that does not overlap with any other sandboxes' virtual memory range. Moreover, this mechanism allows the creation of multiple snapshots of the same function using different virtual memory ranges as a way to reintroduce Address space layout randomization (ASLR). This way, when restoring a function snapshot, the platform could randomly decide which snapshot to use to load the same function into different memory ranges.

The same seed value is also used separate file descriptor and thread id ranges between multiple snapshots. In particular, this ensures that a after restoring a snapshot, the function code will not mistakenly use a file descriptors from another function.


\subsubsection{Ignored system calls and limitations.}
\SYSNAME records and replays system calls that modify system resources. System calls that do not modify the system state such as \texttt{getpid} and \texttt{sched\_yield} are not monitored. Other system calls that change the state of the system are either recorded (\texttt{mmap}, \texttt{open}, \texttt{clone}, etc), or its state is reproduced. For example, \texttt{read} and \texttt{write} are not recorded, but advances in a file-based file descriptor offset can be reproduced by moving the file offset upon a restore operation.

\SYSNAME's snapshots currently do not try to repair network connections. From our experience, language engines quickly reconnect with remote endpoints after a restore. Our current prototype currently also does not checkpoint sub-processes, which could be checkpointed by integrating CRIU~\cite{criu}. We have not found examples of serverless applications that left background processes after an invocation. 
We assume that files produced by the function code will be transported to the restore destination and will not have been modified. 

\subsection{Functions with Native Extensions}
\label{sec:graalvisor:process}
Serverless functions often invoke native extensions in the form of native libraries. These native extensions, commonly used in machine learning inference and video processing workloads for example, extensively use advanced hardware features (such as SIMD and cryptographic instructions) that are not supported by the managed language runtime where the function is running. 
Since native extensions run outside memory isolate sandboxes, where SFI no longer applies, \SYSNAME supports creating sandboxes with stronger isolation by handling function invocations in sub-processes. Such sandboxes are necessary to isolate functions that execute native code, i.e., code that executes outside the sandbox and that may not be ready to receive multiple concurrent invocations. For instance, multiple function invocations executing Python code can have their execution properly contained using sandboxes as the language engine controls in which memory heap objects are allocated. If a particular thread invokes a native method, the isolation guarantees enforced by the engine no longer apply and the thread may be able to access memory that belongs to other functions or corrupt its own managed memory. Furthermore, native libraries may also contain state that is not properly isolated across concurrent function invocations. \SYSNAME can also restore function snapshots with native extensions in a sub-process, allowing functions to benefit from a faster cold start by skipping the initialization of the sandbox entirely in the sub-process. Deciding if a function should execute in a forked sandbox is outside the scope of this work, but could be achieved by analyzing the function code to identify native dependencies at compile or deployment time.

Functions running in sub-process sandboxes are initially loaded into the parent \SYSNAME process. This ensures that both the code cache and the initial heap state are shared in a copy-on-write manner with all sandboxes deployed in child processes. Upon creating a new sandbox, \SYSNAME invokes the Native Sandbox Interface (NSI in Figure~\ref{fig:graalvisor-arch}); if process-isolation is enabled, a sub-process is created. To communicate between the parent and child processes \SYSNAME uses two pipes (returned by the \texttt{pipe} system call): one to send invocation requests from the parent to the child, and one to receive the invocation result from the child to the parent. Both pipes are created right before forking. Upon creating a new forked sandbox, the child process closes parent file descriptors (e.g., the file descriptor used to receive REST requests) and resets signal handlers. Then it creates a new memory sandbox and finally blocks the reading pipe waiting for arguments for a function invocation. The parent side simply creates a process handle with references to both pipes.



\section{High-Density Serverless Platform}
\label{sec:platform}

\begin{figure}[t]
  \centering
  \includegraphics[width=\linewidth]{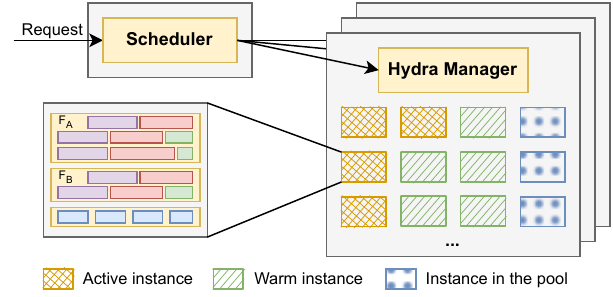}
  \caption{\SYSNAME serverless platform architecture.}
  \label{fig:platform-arch}
\end{figure}

In addition to the proposed runtime, \SYSNAME features a surrounding serverless platform with colocation-aware scheduling policies to maximize deployment density. The top-level component of our platform is the scheduler, which acts as an entry point for the platform. The scheduler controls nodes running \SYSNAME Managers, which in turn manage \SYSNAME runtimes capable of executing the actual function code. \SYSNAME managers maintain a pool of fresh \SYSNAME instances ready to host any function to avoid initializing the virtualization stack on the cold start critical path. Figure~\ref{fig:platform-arch} shows the high-level architecture of the serverless platform.

\subsection{\SYSNAME Scheduler}
\label{sec:platform:scheduler}

To fully leverage \SYSNAME runtime colocation capabilities, we propose a scheduler featuring additional policies to maximize function density. The scheduler governs a fixed number of nodes (\SYSNAME Managers) and keeps track of the state of each node to make scheduling decisions and balance the load. The state of a node includes the node's registered tenants and functions, and estimated memory utilization. In \SYSNAME, functions can have different sizes, determining the maximum memory that can be allocated for this function's sandbox. \SYSNAME scheduler estimates the memory utilization of the node by aggregating the sizes of the functions currently running invocations. Based on the state of the nodes, the scheduler attempts to group invocations in the same nodes (node capacity permitting) using functions and then tenants as a criterion for colocation. 
Finally, if none of the nodes already have the function being invoked or any other function from the same tenant registered, the scheduler selects the least utilized node to serve the invocation.

The scheduler supports four colocation modes:
i) full colocation (all invocations from all tenants can be colocated in the same \SYSNAME runtime);
ii) tenant-based colocation (only functions of the same tenant can be colocated, default for \SYSNAME);
iii) single-function colocation (only invocations of the same function can be colocated);
iv) and no colocation, the execution mode offered by most commercial and open-source serverless platforms. Existing schedulers have no incentive to explore modes i) and ii) because state-of-the-art runtimes do not colocate different functions within a single runtime. Runtimes like Knative~\cite{knative} mostly benefit from mode iii) since Knative supports concurrent invocations to the same function within a single runtime. Mode iv) is suitable for OpenWhisk~\cite{openwhisk} since OpenWhisk runtimes do not support colocation.

\subsection{\SYSNAME Manager}
\label{sec:platform:nodemanager}

The \SYSNAME Manager runs \SYSNAME instances in a local node. It is responsible for managing the lifetime of \SYSNAME instances and scheduling invocations across those instances. When scheduling invocations, the \SYSNAME Manager prioritizes colocating them in a single \SYSNAME instance to benefit from sharing runtime components and avoiding cold starts according to the colocation mode configured in the upstream scheduler, described in~\cref{sec:platform:scheduler}. \SYSNAME instances running invocations in a node are flexible, i.e., they can grow to accommodate parallel invocations and shrink to return memory to the OS. After invocation execution finishes, the \SYSNAME Manager keeps the \SYSNAME instance alive for a configurable amount of time to keep it warm for a potential subsequent invocation.

\SYSNAME runtime is function- and language-generic, enabling \SYSNAME platform to maintain a pool of \SYSNAME instances capable of hosting any function. Instead of creating instances on demand, \SYSNAME Manager fetches a pre-created instance from this pool. This approach avoids initializing the \SYSNAME instance and VM/container on the cold start critical path. As discussed in~\cref{subsec:background:impact}, with runtimes like OpenWhisk~\cite{openwhisk}, platforms can also maintain a pool of instances. However, OpenWhisk runtime instances are language-specific, implying that platforms would have to maintain separate pools for each language, whereas the \SYSNAME platform only needs to maintain a single pool without speculatively partitioning it between languages. For Knative, platforms cannot pre-create instances since Knative instances are function-specific. Besides, the \SYSNAME platform does not have to resize \SYSNAME instances when registering a function since \SYSNAME runtime can create memory-constrained sandboxes (factor \#2 in~\cref{subsec:background:impact}). In addition, \SYSNAME runtime can destroy sandboxes after a configurable period of inactivity, effectively returning memory to the OS, unlike Knative, where concurrent requests share the same heap that cannot shrink (factor \#1 in~\cref{subsec:background:impact}).

A periodic background job monitors the state of the pool to prevent it from complete depletion. If the pool is depleted to a configurable threshold during execution, this job reclaims a configurable number of least-recently-used \SYSNAME instances to replenish the pool with fresh instances. The rest of the warm \SYSNAME instances remain untouched and ready to serve future invocations.

\section{Evaluation} \label{sec:evaluation}




This section analyzes the performance of \SYSNAME and compares it to other state-of-the-art runtimes and platforms. \SYSNAME is implemented in a combination of Java and C in 10298 lines of code which we plan to make available to public. We start by measuring the density of different runtimes using a variety of serverless benchmarks (\cref{sec:evaluation:tputmem}). Then, we replay a real serverless invocation trace on a local \SYSNAME platform deployment (\cref{sec:evaluation:cluster}) to evaluate \SYSNAME at the platform level. Table~\ref{table:benchmarks} lists the benchmarks used in our experiments. We combine benchmarks from Sebs~\cite{copik:2021} (Python and JavaScript benchmarks) and Photons~\cite{dukic:2020} (Java benchmarks). 

\begin{table}
\centering
\begin{tabular}{l|p{5.5cm}}
\hline
 \textbf{Language} & \textbf{Benchmark} \\ \hline
 JavaScript (js) & helloworld (hw), dynamic-html (html), upload \\ \hline
 Python (py) &  helloworld (hw), dynamic-html (html), thumbnail, upload, compress, video-processing (video), minimum spanning tree (mst), breadth-first search (bfs), pagerank (pr), dna \\ \hline
 Java (jv) &  helloworld (hw), filehashing (hash), restapi (rest), video-processing (video), classify \\ \hline
\end{tabular}
\caption{List of benchmarks per language.}
\label{table:benchmarks}
\end{table}

\subsection{Function Sandbox Density} \label{sec:evaluation:tputmem}
\begin{figure*}[t]
  \begin{minipage}[t]{\linewidth}
    \centering
    \includegraphics[width=.95\linewidth]{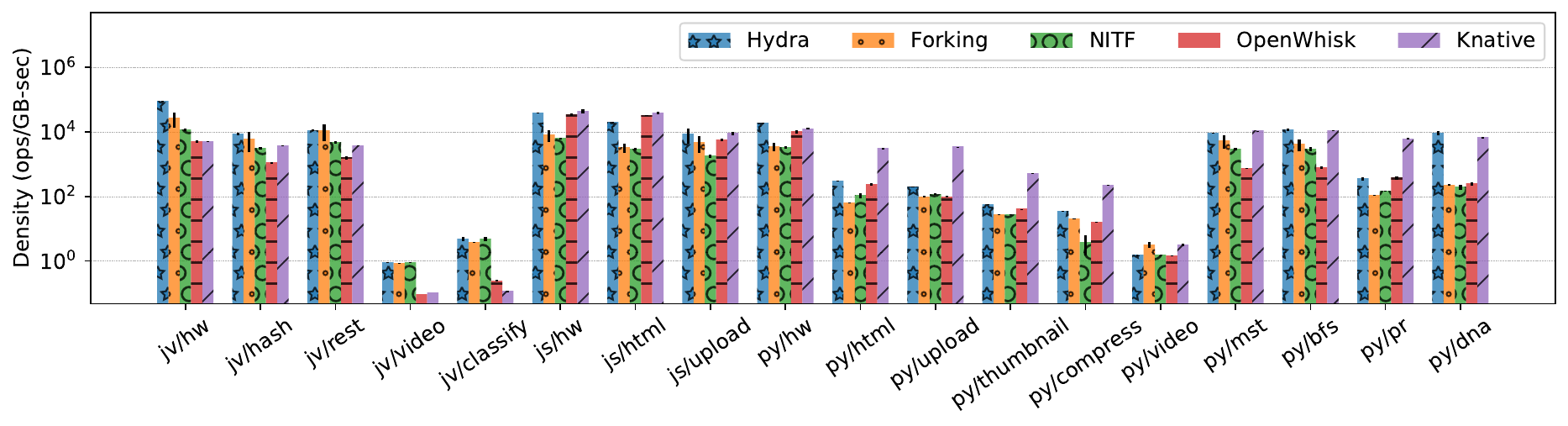}
  \end{minipage}
  \caption{Density of different serverless runtimes for different benchmarks.}
  \label{fig:eval:sandboxes:efficiency}  
\end{figure*}

\begin{figure}[t]
  \centering
  \includegraphics[width=\linewidth]{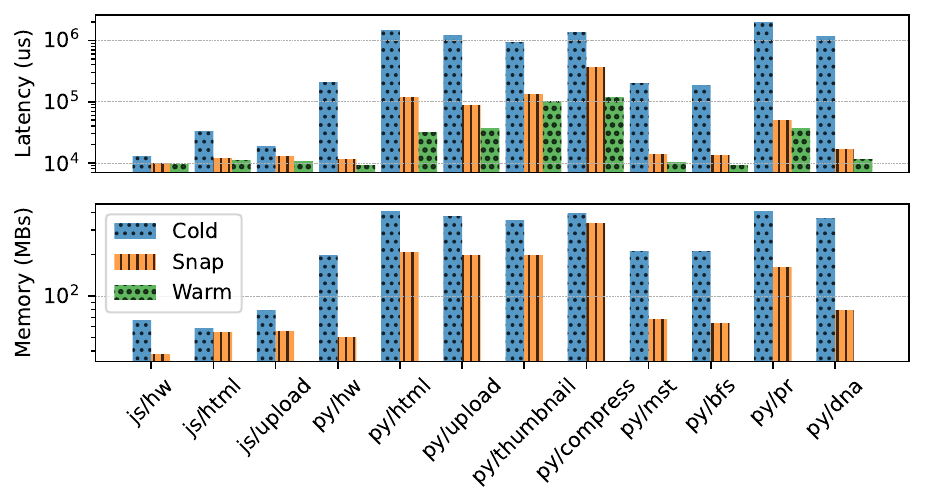}
  \caption{Invocation latency and memory footprint comparing a new sandbox (Cold) to a restored from snapshot (Snap) and warmed-up (Warm) in \SYSNAME.}
  \label{fig:eval-snapshot}
\end{figure}

This experiment analyzes the function sandbox density of four different types of serverless runtimes:
\begin{itemize}[leftmargin=*]
    \item \textbf{OpenWhisk}~\cite{openwhisk} runtimes used to handle Java (JVM), JavaScript (Node.js), and Python (CPython) invocations. These runtimes are state-of-the-art, native language implementations. Each of these runtimes can handle invocations of a single function, one invocation at a time;
    \item \textbf{Knative}~\cite{knative} runtimes are similar to OpenWhisk but capable of handling multiple concurrent invocations of the same function. Parallel invocations are not isolated from each other. Another difference is that it uses GraalVM Native Image~\cite{graalvm} to compile Java functions into native executables. For Python and JavaScript, Knative uses CPython and Node.js by default, respectively;
    \item \textbf{NITF}, a variant of \SYSNAME that does not allow multiple functions to execute in the same instance. This baseline mimics what is possible to achieve today by combining GraalVM Native Image and Truffle. Similarly to OpenWhisk runtimes, this variant can handle invocations of a single function, one invocation at a time;
    \item \textbf{Forking}, a variant of \SYSNAME inspired by SAND~\cite{akkus:2018}. In this system, multiple invocations of a single function can be served concurrently. Each invocation will be handled in a fork of the parent process that pre-loaded the function code. We use this runtime to compare the overheads of process forking compared to sandboxes;
    \item \textbf{\SYSNAME}, capable of handling multiple function invocations from different functions at the same time. We allow \SYSNAME to use already snapshotted sandboxes (see Figure~\ref{fig:eval-snapshot} where we measure the effect of snapshotting).
\end{itemize}

Our goal is to measure how much throughput can be produced on a single core with 2 GBs of memory~\footnote{The used ratio is inspired by the 1:769 core-to-memory fixed ratio used in Amazon Lambda~\cite{lambdaquotas}, which we approximate to 1:2 for simplicity.}. Our target metric, density, is defined as the ratio of throughput per memory (ops/GB-sec). Naturally, higher density can be achieved through higher throughput or lower memory consumption. Since several benchmarks are I/O-bound (e.g., \texttt{jv/hash}, \texttt{js/uploader}, \texttt{py/hw}), we allow concurrent requests to share the virtualization stack to improve density. To do so, we add additional concurrent requests until the density of the runtime stops increasing. For OpenWhisk and NITF, handling multiple requests concurrently means having several instances running with a fraction of a core and a fraction of 2 GBs. For \SYSNAME and Forking, we use a single instance with multiple sandboxes. 


Experiments are conducted on a single cluster node running Linux kernel 5.15.0 equipped with 2x Intel Xeon Gold 5320 CPU @ 2.2GHz and 128GB of DDR4 DRAM. CPU frequency scaling and hyper-threading are disabled, and runtimes are pinned to a single core. Each runtime runs inside a Docker container with limited CPU and memory. Throughput is measured using Apache Bench,\footnote{https://httpd.apache.org/docs/2.4/programs/ab.html} and memory represents sum of the RSS of the runtime (and children processes if any). The results are presented in Figure~\ref{fig:eval:sandboxes:efficiency}. Each bar represents an average of five iterations, where each iteration includes 10-16000 requests depending on the duration of each function invocation (which ranges from a few milliseconds to a few seconds).

\subsubsection{Java Benchmarks}
\label{sec:eval:java}
\SYSNAME presents 14$\times$ higher density compared to OpenWhisk and 8.6$\times$ higher efficiency compared to Knative. Compared to OpenWhisk, for Java IO-bound benchmarks (\texttt{hw}, \texttt{hash}, and \texttt{rest}), \SYSNAME reduces the number of runtime instances to one, therefore significantly reducing the memory footprint. In addition, CPU multiplexing improves by having a full core shared across several sandboxes compared to having multiple quotas of a core for each OpenWhisk container. 
Compared to Knative, \SYSNAME has 3$\times$ higher average memory footprint in IO-bound benchmarks since unlike Knative, \SYSNAME has separate heaps and other language runtime components for each concurrent invocation. However, \SYSNAME has 10.9$\times$ higher throughput on average due to a more lightweight server stack compared to Knative, which by default uses Spring Cloud Functions~\cite{spring-functions}.
Forking has the closest performance compared to \SYSNAME. The slight density degradation comes mostly from frequent context switching and the increased memory footprint that comes from duplicated pages in the child process. 


\subsubsection{JavaScript and Python Benchmarks}
\label{sec:eval:poly}
For JavaScript and Python benchmarks, \SYSNAME improves the average density by 1.4$\times$ compared to OpenWhisk. It allows multiple sandboxes to share the same Truffle runtime components and benefits from sandbox snapshotting to bypass Truffle context initialization overheads. In several benchmarks, however, OpenWhisk registers higher density. After analyzing these runs we conclude that these results stem from performance issues in the implementation of Truffle engines which are still maturing compared to Node.js and CPython. This leads to lower throughput and higher memory footprint in particular in \texttt{js/html}, \texttt{py/video}, and \texttt{py/pagerank}. \SYSNAME shows 1.2$\times$ lower average efficiency compared to Knative, which is capable of handling concurrent invocations. However, Knative does not provide isolation within a single runtime and forces developers to separate the state of concurrent invocations, while \SYSNAME isolates concurrent invocations in sandboxes and scales transparently following the principles of serverless.

\subsubsection{Function Snapshotting Impact on Density}
\label{sec:eval:snapshot}
We now evaluate the impact of sandbox snapshotting in \SYSNAME by comparing the latency and memory footprint required to handle a cold start (launching a new sandbox and invoking the function). Note that for sandboxes created from scratch, this may require code interpretation and JIT-compilation. 

Results are depicted in Figure~\ref{fig:eval-snapshot}. Both plots include Cold (vanilla sandboxes that suffer from a cold start) and Snap (restored sandboxes that restore the execution past a cold start), and the latency plot includes Warm (restored sandboxes that warmed up with five requests). The memory footprint does not change after restore (Snap); thus, we do not report Warm in the memory plot. We omit benchmarks that execute in a forked process (e.g., \texttt{jv/video} and \texttt{py/video}), something that our C/R mechanism is not designed to support (\cref{sec:graalvisor:cr}). Besides, we omit Java benchmarks as GraalVM Native Image already offers fast startup for Java applications, and snapshotting offers minimal benefits for both latency and memory footprint. Note the log scale on both y-axes. Overall, snapshotting reduces the latency to serve the first request by 16.5$\times$ and reduces the memory footprint by 2.4$\times$.

\subsubsection{Function Checkpointing Overhead}
\label{sec:eval:checkpoint-overhead}
Checkpointing the function during its execution induces additional overheads described in~\cref{sec:graalvisor:cr}. Using the benchmarks from Table~\ref{table:benchmarks}, we observe that the function execution that produces a snapshot (i.e., involves collecting and analyzing system calls, etc.) is on average 7.1$\times$ slower compared to a cold start not involving checkpointing routines. However, checkpointing needs to be executed only once for each function, and its costs are amortized with the subsequent invocations to this function.

\subsection{Reproducing a Serverless Trace}
\label{sec:evaluation:cluster}

So far, we have analyzed the performance of different serverless runtimes when executing multiple invocations a single function. However, in real environments, platforms deal with concurrent requests from different functions. To implement such an environment, we take advantage of the public Azure Functions trace~\cite{sharad:2020}. The trace provides data about function invocations, including the function identifier, duration, memory footprint, and the tenant associated with the function.

\subsubsection{Top-Level Scheduling Setup}

\begin{figure}[t]
  \centering
  \includegraphics[width=.65\linewidth]{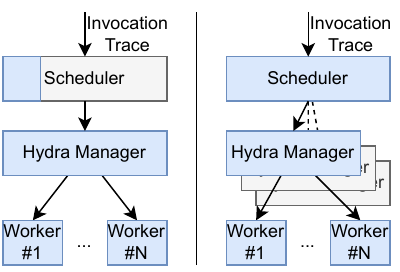}
  \caption{Serverless traces, downscaled with random or statistic-driven approaches to 25\% of the original trace, will only use 25\% of the platform's control plane, whereas running the full trace will use 100\% of the platform's control plane. Blue color represents actively used capacity, and gray represents unused capacity.}
  \label{fig:downsampling}
\end{figure}

\begin{figure}[t]
  \centering
  \includegraphics[width=\linewidth]{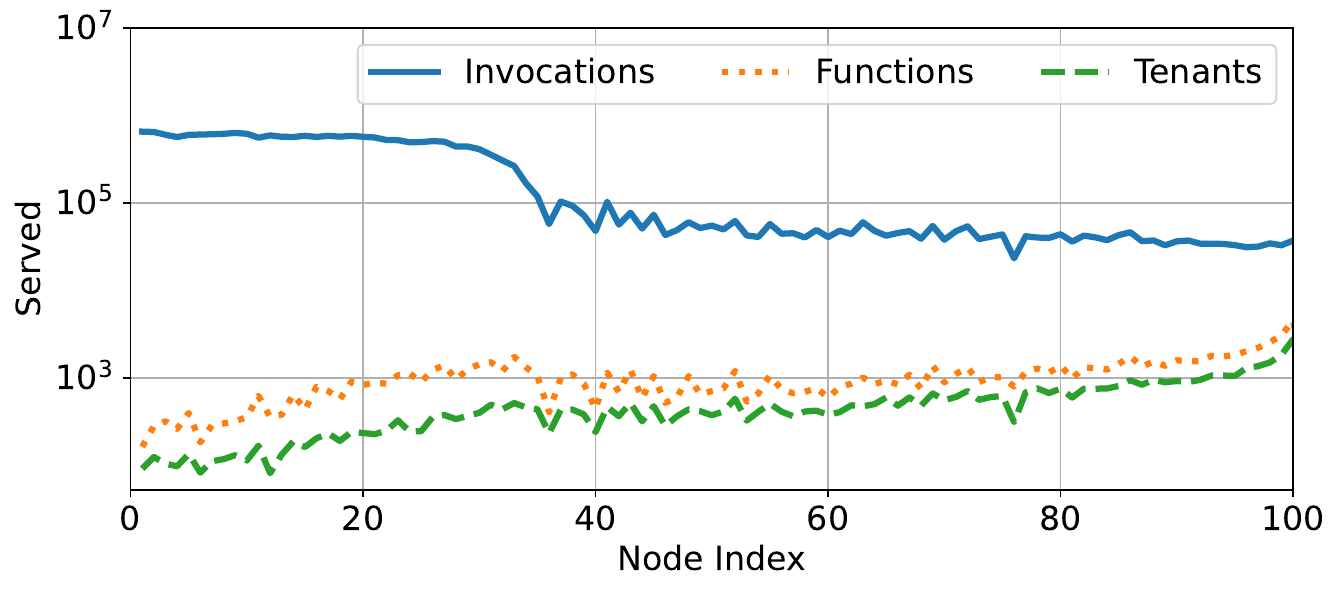}
  \caption{Distribution of requests, registered functions and tenants across nodes in the platform.}
  \label{fig:azure:nodes}
\end{figure}

Existing approaches to evaluating the impact of virtualization stacks in a realistic serverless scenario often involve random or statistic-driven down-sampling~\cite{ustiugov:2023,zhang:2023,singhvi:2021,miao:2024,fuerst:2021} of the real-world serverless traces to run workloads on a lower scale. While these down-sampling techniques allow pushing the load to the desired portion of the serverless platform's data plane (i.e., invocations served in a particular node or worker), the load on the control plane (i.e., invocation scheduling) is partially omitted (see Figure~\ref{fig:downsampling}, left side). Therefore, we argue that such approaches are applicable at the node or worker level but do not account for the platform scheduling that, depending on its policies, may impact the way virtualization stacks are utilized. Besides, random or statistic-driven down-sampling may require multiple experimental iterations to select a sample trace for a particular testbed, as they do not adapt with back-pressure when the node struggles with a particular set of requests.

The scheduler described in~\cref{sec:platform:scheduler}, in addition to running workloads in production, can also be used as a benchmarking tool for the serverless platform. In an experimental setup, the scheduler runs the original (full) trace. The nodes (some portion or even all of them) can simulate serving the invocations instead of running the actual functions, thus not requiring computational resources. This way, the control plane is loaded realistically since the scheduler processes the entire original trace (see Figure~\ref{fig:downsampling}, right side). Besides, the scheduler adapts to back-pressure that prevents it from overloading nodes since the scheduler keeps track of the state of each node.


Analyzing the Azure Function trace~\cite{sharad:2020}, it is possible to demonstrate that a minority of functions generates the vast majority of all invocations. Therefore, with colocation-oriented scheduling, the distribution of invocations and registered functions across the nodes of the serverless platform can be nonuniform, as some functions generate more load than others. Figure~\ref{fig:azure:nodes} demonstrates that some nodes tend to have few functions registered, albeit these functions get more invocations than others. Other nodes observe the opposite situation: they tend to have more different functions registered, but these functions are invoked infrequently. With this insight, we sample the performance of the \SYSNAME platform in two different nodes, one with low heterogeneity (invocations dominated by a few functions, labeled Node \#1) and another with high heterogeneity (invocations originate from many different functions, labeled Node~\#2).

\begin{figure}[t]
  \begin{subfigure}{\linewidth}
    \includegraphics[width=\linewidth]{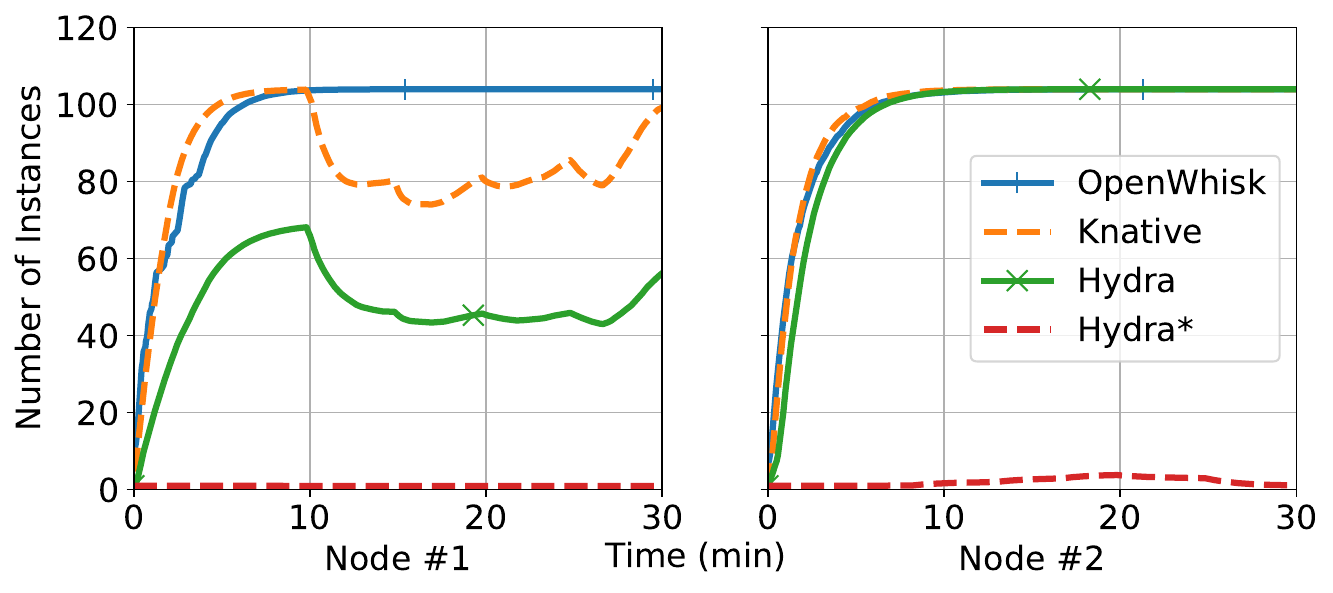}
    \caption{Number of active runtimes.}
  \end{subfigure}
  \begin{subfigure}{\linewidth}
    \includegraphics[width=\linewidth]{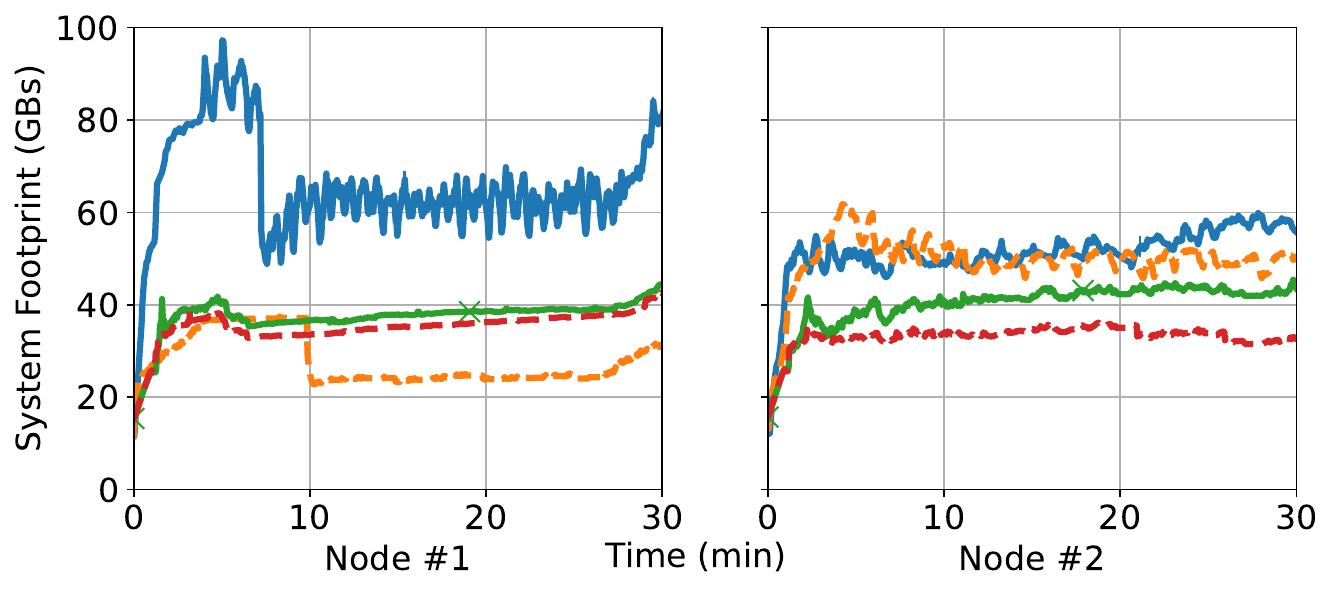}
    \caption{System memory footprint.}
  \end{subfigure}
   \begin{subfigure}{\linewidth}
    \includegraphics[width=\linewidth]{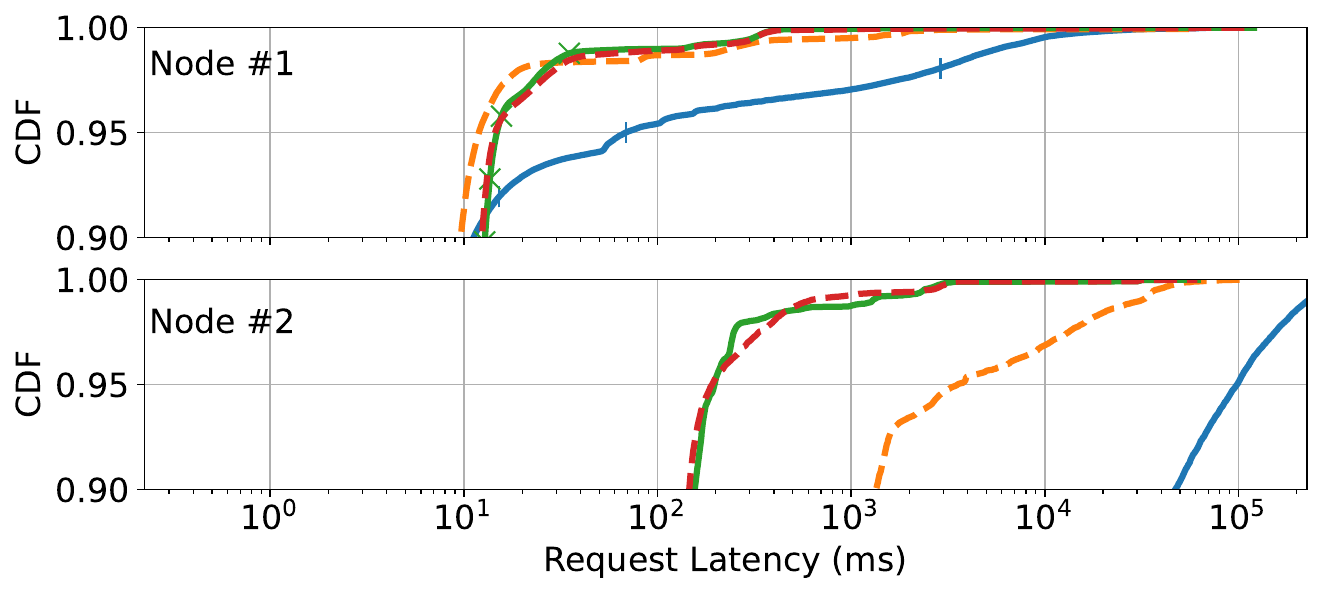}
    \caption{End-to-end user request latency.}
  \end{subfigure}
  \caption{Replaying a serverless invocation trace from Azure~\cite{sharad:2020} on different serverless runtimes. Node \#1 served 167043 invocations of 157 functions from 86 tenants. Node \#2 served 23267 invocations of 793 functions from 311 tenants.}
  \label{fig:azure:replay}
\end{figure}

\subsubsection{Reproducing a Real-World Trace}

We select a 30-minute window that represents the average invocation activity in serverless. We analyzed all other 30-minute windows throughout the same day of the trace to ensure that the selected trace exhibits the same common profile in terms of the number of invocations processed, and total functions and tenants registered. We reproduce the selected segment on a local cluster machine with the same specifications as in~\cref{sec:evaluation:tputmem}. The scheduler runs on another machine with identical specifications. As the function workload, we use the benchmarks from Table~\ref{table:benchmarks}. Each benchmark is configured to use limited resources according to its complexity, 1/8 CPU and 256 MB of memory being the minimal resource allocation for the function, and 1 CPU and 2048 MB being the maximum resource allocation. We assign languages to functions from the trace according to the recent study on runtimes in serverless~\cite{newrelic} to maintain a realistic language distribution, and assign benchmarks (workloads) to functions based on the average duration of each function from the trace.


In this experiment, we replay the same serverless invocation trace from the scheduler machine with four execution modes: \SYSNAMEPRIME (colocating all invocations), \SYSNAME (colocating invocations to functions of the same tenant), Knative~\cite{knative} (colocating invocations of the same function), and OpenWhisk (no colocation). The scheduler operates with 99 nodes that simulate invocations and one node that executes the actual function code. The \SYSNAME platform maintains a pool of 416 pre-created \SYSNAME workers (containers running \SYSNAME runtimes) as described in \cref{sec:platform:nodemanager}. The total memory footprint of this pool is 1.9 GB. For Knative and OpenWhisk, workers are created on demand since for OpenWhisk, the pool would have to be partitioned across languages, and the platform cannot resize workers (language runtimes) after they are instantiated, and for Knative, workers are function-specific (as discussed in~\cref{subsec:background:impact}). For all modes, the platform does not run more than 416 concurrent workers and reclaims the least recently used ones when it needs to create more. The platform keeps each worker alive for a randomized time period between 10 and 20 minutes after the last invocation.

\begin{table}
  \centering
  \begin{subtable}{0.45\linewidth}
    \centering
    \begin{tabular}{l||c|c}
    \hline
     \textbf{Mode} & \textbf{Cold} & \textbf{Warm}\\ \hline
     OpenWhisk & 2821 & 0\\ \hline
     Knative & 209 & 0\\ \hline
     Hydra & 0 & 118\\ \hline
     Hydra* & 0 & 1\\ \hline
    \end{tabular}
    \caption{Node \#1 (167K invocations).}
  \end{subtable}
  \hspace{2em}
  \begin{subtable}{0.45\linewidth}
    \centering
    \begin{tabular}{c|c}
    \hline
     \textbf{Cold} & \textbf{Warm}\\ \hline
     3049 & 0\\ \hline
     2064 & 0\\ \hline
     0 & 408\\ \hline
     0 & 4\\ \hline
    \end{tabular}
    \caption{Node \#2 (23K invocations).}
  \end{subtable}
\caption{Cold and warm starts from replaying the serverless invocation trace.}
\label{table:starts}
\end{table}

The results in Figure~\ref{fig:azure:replay} include the number of active instances (a), the total memory consumption of the node (b), and the user request latency (c). In addition, Table~\ref{table:starts} shows the number of warm (fetching a worker from the pool) and cold (launching a new worker) starts. The "Node \#1" label denotes the node serving a few frequently invoked functions, and the "Node \#2" label denotes the node with many infrequently invoked functions.

Overall, by being able to consolidate invocations of different functions in \SYSNAME instances, it significantly reduces the number of active instances without degrading latency, leading to memory reduction of up to 21.3-40.4\% on average compared to OpenWhisk in both nodes if using \SYSNAME, and up to 35.6-43.9\% if using \SYSNAMEPRIME. In Node \#1, Knative shows lower memory footprint compared to \SYSNAME modes because only a few functions generate most of the load, and only one Knative worker is needed for each function. However, it is important to note that the lower memory footprint offered by Knative comes as a consequence of the lack of isolation between concurrent requests within each worker, meaning that the heap and the other runtime components will be shared. \SYSNAME, on the other hand, runs concurrent invocations in sandboxes that do not share runtime components, leading to a higher memory footprint. In Node \#2, \SYSNAME reduces the memory footprint by 14.5\%, and \SYSNAMEPRIME reduces the memory footprint by 30\% compared to Knative because Knative needs one worker per function, whereas \SYSNAME colocates functions from the same tenant, and \SYSNAMEPRIME colocates all functions.

\SYSNAME reduces tail latency by 51.4-375.5~$\times$ (p99) in Node \#2 compared to Knative and OpenWhisk, respectively (Figure~\ref{fig:azure:replay}.c), as it fetches pre-created \SYSNAME workers from the pool instead of initializing them on demand. OpenWhisk and Knative, on the other hand, experience numerous cold starts (see Table~\ref{table:starts}.b) for many different functions invoked infrequently. In Node \#1, where a few functions receive the majority of invocations, the impact of cold starts for Knative and, to a lesser extent, for OpenWhisk is amortized by the high number of invocations to the few functions that can reuse workers. Still, \SYSNAME reduces p99 latency by 45.3~$\times$ compared to OpenWhisk since OpenWhisk needs to initialize and run more workers to accommodate concurrent requests, and by 1.9~$\times$ compared to Knative since Knative cannot colocate different functions even from the same tenant.

\textbf{Summary.} By combining a virtualized runtime, colocation-aware scheduler, and a local pool of pre-created generic runtime instances, \SYSNAME significantly increases density, reduces memory footprint, and eliminates cold starts, thus reducing tail latency.



\section{Related Work} \label{sec:relatedwork}
\noindent\textbf{Runtime Forking and Snapshotting} have been widely studied in the context of serverless. SAND~\cite{oakes:2018} and SOCK~\cite{akkus:2018} rely on process forking to minimize cold start invocation latency and reduce memory footprint by sharing runtime libraries. Other systems such as FaaSnap~\cite{ao:2022}, SEUSS~\cite{cadden:2020}, Catalyzer~\cite{du:2020}, vHive~\cite{ustiugov:2021}, Prebaking~\cite{silva:2020}, Fireworks~\cite{shin:2022}, Medes~\cite{saxena:2022}, and Sabre~\cite{lazarev:2024} proposed different snapshotting techniques to reduce cold start latency and memory footprint. Snapshotting alone, however, does not reduce resource redundancy as multiple virtualization stacks are still launched from a common snapshot.

\SYSNAME aggregates multiple function invocations in lightweight sandboxes in a single virtualization stack. Doing so allows applications to use process-level isolation but can also support in-process memory sandboxes, which are faster to launch and lead to higher density (\cref{sec:evaluation:tputmem}). \SYSNAME can additionally snapshot individual sandboxes 
(\cref{sec:graalvisor:cr}) to reduce the cold start latency of a new \SYSNAME function sandbox. Deciding when to checkpoint a particular sandbox is outside the scope of this work and solutions proposed in previous works (such as Pronghorn~\cite{kohli:2024}) can be utilized.

\noindent\textbf{Runtime Pre-Warmup} has also been studied with the goal of launching runtimes just before invocations are received~\cite{mahgoub:2022} and thus hiding the cold start latency. Others have studied how to reduce the pre-warmup cost by taking advantage of heterogeneous resources~\cite{roy:2022}. Mohan et al.~\cite{mohan:2019} proposed a pool of containers that could be utilized when function invocations arrive. TrEnv~\cite{huang:2024} offered a pool of repurposable container resources that could be used with any function. Our platform pre-warmups entire \SYSNAME instances (container and runtime) in the pool as described in~\cref{sec:platform:nodemanager} but could also be integrated with existing works that predict future function invocations. 

However, replacing multiple runtimes that implement a single language with a single runtime that implements multiple languages also makes it easier to manage runtime images. By doing so, predicting which runtime to use for the next invocation becomes a lesser problem. A pool of \SYSNAME runtimes can be used to accommodate any function invocation supported by \SYSNAME. 

\noindent\textbf{Virtualizing Runtimes} allows one or more functions to share a single runtime. Sharing, however, can be achieved with different levels of isolation. Crucial~\cite{pons:2019} and Nightcore~\cite{jia:2021} handle multiple function invocations in a single runtime using different threads. Besides, Nightcore uses forking to spawn runtime processes, effectively eliminating support for such languages as Java. Photons~\cite{dukic:2020} also runs concurrent function invocations in a single runtime but relies on Java-specific bytecode transformations to ensure isolation (i.e., concurrent function invocations could not interact with other co-executing invocations). Boucher et al.~\cite{boucher:2018} rely on Rust's memory safety guarantees to handle multiple invocations concurrently in isolation. Cloudflare Workers~\cite{cloudflare} rely on V8 memory isolates to run concurrent JavaScript functions in the same V8 runtime instance. Faasm~\cite{shillaker:2020,segarra:2025}, Sledge~\cite{gadepalli:2020}, and Gackstatter et al.~\cite{gackstatter:2022} proposed using WebAssembly~\cite{haas:2017} to execute multiple invocations in isolation. WebAssembly, however, still presents some challenges when running high-level languages such as Java, JavaScript, and Python (see~\cref{subsec:exec:environments} for more details). As a result, runtimes such as Faasm need to compile a Python runtime such as CPython to WebAssembly in order to run Python functions. If concurrent invocations are to be handled, multiple runtimes will run concurrently in different WebAssembly execution environments, leading to resource redundancy.

\SYSNAME enforces isolation through memory sandboxes. However, compared to Cloudflare Workers, \SYSNAME is not limited to JavaScript or WebAssembly functions. \SYSNAME also supports applications that carry native dependencies through forked sandboxes and also supports sandbox snapshotting. 
Opposed to Faasm (which is centered on WebAssembly-based functions), \SYSNAME relies on Truffle to support multiple language support, and therefore, concurrent invocations always share the underlying runtime. In addition, to further share the JIT compilation and profiling, \SYSNAME shares the code caches among concurrent invocations of the same function. 



\noindent\textbf{Profile and Code Cache Sharing} reduce code profiling and compilation overheads. Profile sharing has been proposed by Arnold et al.~\cite{arnold:2005} and Ottoni et al.~\cite{ottoni:2021} as a way to re-use profiling information. By doing so, future invocations do not need to repeat all the code profiling steps and can benefit from the profiles gathered by previous invocations. A similar technique has also been used to share Just-In-Time compiled code \cite{khrabrov:2022,xu:2018} among different invocations of the same application/function. Such techniques could be integrated into \SYSNAME to share Truffle code caches. However, solutions such as JITServer~\cite{khrabrov:2022} and ShareJIT~\cite{xu:2018} are designed to optimize long-running applications and are not prepared to deal with the challenges inherent to fast-executing and lightweight serverless functions~\cite{carreira:2021}.

\section{Conclusions}

\SYSNAME offers a virtualized multi-language runtime capable of colocating different functions running in lightweight isolated sandboxes sharing the same virtualization stack. To take advantage of the runtime, \SYSNAME features a colocation-aware scheduler and a local \SYSNAME Manager that maintains a pool of pre-created \SYSNAME instances. As a result, \SYSNAME eliminates cold starts when executing the realistic serverless trace (reducing invocation tail latency), reduces memory footprint, and improves density compared to state-of-the-art serverless runtimes. We plan to make the \SYSNAME project sources available to the public.


\bibliographystyle{plain}
\bibliography{references}

\end{document}